\documentclass[11pt, a4paper]{article}

\usepackage{graphicx}
\usepackage{bm}
\usepackage{amsmath, amssymb,theorem,verbatim}
\usepackage[round]{natbib} 
\usepackage{setspace}
\usepackage{multirow} 
\usepackage{hyperref}
\usepackage{caption}
\usepackage{subcaption}
\usepackage{algpseudocode}
\usepackage{textcomp}

\long\def\comment#1{}

\newtheorem{theorem}{Theorem}[section]

\newtheorem{prop}{Proposition}[section]

\newtheorem{defin}{Definition}[section]

\newtheorem{corollary}{Corollary}[section]

\newtheorem{assumption}{Assumption}[section]

\begin{document}


\author{Piotr Fryzlewicz\thanks{Department of Statistics, London School of Economics, Houghton Street, London WC2A 2AE, UK. Email: \url{p.fryzlewicz@lse.ac.uk}.
Work supported by the Engineering and Physical Sciences Research Council grant no.
EP/L014246/1.}}

\title{Detecting possibly frequent change-points: Wild Binary Segmentation 2 and steepest-drop model selection}

\oddsidemargin=0.25in
\evensidemargin=0in
\textwidth=6in
\headheight=0pt
\headsep=0pt
\topmargin=0in
\textheight=9in

\maketitle

\begin{abstract}
Many existing procedures for detecting multiple change-points in data sequences fail
in frequent-change-point scenarios. This article proposes a new change-point
detection methodology designed to work well in both infrequent and frequent
change-point settings. It is made up of two ingredients: one is ``Wild Binary
Segmentation 2'' (WBS2), a recursive algorithm for producing
what we call a `complete' solution path to the change-point detection problem, i.e.
a sequence of estimated nested models containing $0, \ldots, T-1$ change-points, where
$T$ is the data length.
The other ingredient is a new model selection procedure,
referred to as ``Steepest Drop to Low Levels" (SDLL). The SDLL criterion acts on
the WBS2 solution path, and, unlike many existing model selection procedures for
change-point problems, it is not penalty-based, and only uses thresholding as a certain
discrete secondary
check. The resulting WBS2.SDLL procedure, combining both ingredients, is shown to
be consistent, and to significantly outperform the competition in 
the frequent change-point scenarios tested. WBS2.SDLL is fast, easy to code and
does not require the choice of a window or span parameter.

\vspace{5pt}

\end{abstract}

\section{Introduction}
\label{sec:intro}
A recent report \citep{nap13}
lists change-point detection as one of the ``inferential giants'' (common building blocks in
knowledge discovery) in massive data analysis. Change-point analysis, be it a posteriori or online,
is used in fields as diverse as bioinformatics \citep{ovlw04}, econometrics and finance \citep{h01, ag02, bp03}, climate and the natural
environment \citep{rcwll07, lcll10, saad12, pzbac13}, autonomous driving \citep{gceo15},
computer vision \citep{r12}, 
clinical neuroscience \citep{yam14}
and quality/reliability monitoring \citep{aptlbc11, hl11, aa12}.

A common testing ground for a posteriori change-point detection methods, and 
the focus of this paper, is the model
\begin{equation}
\label{eq:model}
X_t = f_t + \varepsilon_t,\quad t = 1, \ldots, T,
\end{equation}
in which $f_t$ is a deterministic, one-dimensional, piecewise-constant signal with 
change-points whose number $N$ and locations $\eta_1, \ldots, \eta_N$ are unknown. 
The sequence $\varepsilon_t$ is random and such that $\mathbb{E}(\varepsilon_t)$ is 
exactly or approximately zero. In the simplest case $\varepsilon_t$ are modelled as 
i.i.d. $N(0, \sigma^2)$, but can also
have other marginal
distributions
and/or exhibit temporal dependence. The task is to estimate
$N$ and $\eta_1, \ldots, \eta_N$ under various assumptions on $N$, the magnitudes
of the jumps and the minimum permitted distance between the change-point locations.
This article focuses on the i.i.d. $N(0, \sigma^2)$ distribution for $\varepsilon_t$, as even
in this simplest case, the above change-point detection problem still poses major 
challenges for the state of the art, as this article illustrates.

Although performance measures for the above estimation task, naturally, vary across the literature,
it can be anticipated that many users will have a preference for methods which
\begin{enumerate}
\item
accurately estimate the number $N$ and locations $\eta_i$ of any change-points present in
$f_t$, for both
\begin{enumerate}
\item
signals with few/infrequent change-points, and
\item
signals with many/frequent change-points;
\end{enumerate}
\item
are fast to execute and scale well to long signals.
\end{enumerate}
Depending on the setting, other desirable features may include e.g. the existence of software,
 ease of implementation
and transferability to other, more complex stochastic models.
Even though much methodological and algorithmic progress has been made on the above change-point
detection problem in recent years, we show in this paper that very few (if any) state-of-the-art
techniques simultaneously possess the above desired features 1(a), 1(b) and 2.
In particular, many techniques perform surprisingly poorly for signals with 
frequent change-points.
We first briefly describe the history of the multiple change-point detection problem and the state of the
art. For reasons of space, our literature review mainly focuses on the parametric case, i.e. models
in which the distribution of $X_t$ can be modelled parametrically, with special
emphasis on the i.i.d. Gaussian model for $\varepsilon_t$.

A major class of approaches to a posteriori multiple change-point detection is one in which 
change-points are found by minimising
a criterion function consisting of a likelihood-type (or least-squares)
term measuring the fit of the estimate to the data plus a penalty term 
to prevent overfitting. In this category,
\cite{ya89} consider least-squares estimation of $f_t$ for a fixed $N$
and i.i.d. noise.
\cite{bbm00} extend this work to noise for which the variance is a function
of the mean.
In the Gaussian case, the SIC (BIC) is used to estimate an unknown but bounded
$N$ in \cite{y88}, and a more general penalty but also linear in the number of change-points
appears in \cite{l95}. For an unknown but bounded $N$, \cite{lm00} consider penalised least-squares
estimation, with a penalty linear in the number of change-points, for dependent $\varepsilon_t$'s;
see also \cite{l99a} and \cite{l05a}, an article we return to later. For a fixed $N$, \cite{pc06} propose a 
likelihood criterion with a 
penalty depending not only on the number, but also on the locations of change-points, favouring 
more uniformly-spread estimated change-points; related ideas appear in \cite{zs07}.
For an unknown $N$, \cite{l05}
proposes least-squares estimation with a penalty originating from the model selection
approach of \cite{bm01}; we revisit this approach below. \cite{bklmw09} use the least-squares criterion with a linear penalty on the number of 
change-points. More general forms of Schwarz-like penalties are studied, for example, in \cite{w08}, 
\cite{c11} and \cite{c14}. The SMUCE estimator of \cite{fms14} can also be cast in the
penalised cost function framework (while SMUCE is shown to control the FWER, the related method
of \cite{lm16} controls the FDR).
An empirical Bayes approach to change-point estimation in a marginal 
likelihood framework appears in \cite{dkk16}.
The PELT algorithm by \cite{kfe12} and the pruned dynamic programming
by \cite{r10} accelerate computation of some penalised change-point estimators
to linear time in best-case scenarios
(while retaining quadratic speed in worst-case ones),
and offer fast implementations. 
Related ideas appear also in \cite{mhrf14}. 

Other attempts to reduce the computational complexity of the 
problem include \cite{dlry06}, who (in a time series setting) use
a genetic algorithm to minimise a Minimum Description Length criterion, 
and \cite{hll10} who consider the least-squares
criterion with a total variation penalty, which enables them to use the LARS 
algorithm of \cite{ehjt04} to compute the solution in $O(N T \log(T))$ time.
However, the total variation penalty
is not an optimal one for change-point detection (in the sense of balancing out
type-I and type-II errors, as described in \cite{bd93} and \cite{cf11}). The total variation penalty
is also considered in the context of peak/trough detection by \cite{dk01}, 
who propose the `taut string' approach for fast computation. In the context of 
multiple change-point detection, it is considered by \cite{r09} (with a subsequent
correction published on the author's web page) and \cite{rw14} as part
of the fused lasso penalty, proposed by \cite{tsrz05} and equivalent to taut string
in model (\ref{eq:model}).
\cite{lsrt17} propose a generic post-processing procedure for any $L_2$-consistent,
piecewise-constant estimator of $f_t$, including in particular the total-variation-based
one, which yields consistency for the estimators of $\eta_i$ under certain assumptions
on the distances between the change-points and the jump sizes. Similar results for 
general ``trend filtering'' \citep{t14} estimators appear in \cite{glcs18}.

Our experience is that many penalty-based methods in which the penalty has 
been pre-set by the user (rather than having been chosen adaptively from the data)
can struggle to offer uniformly good performance across both signals with infrequent change-points,
and signals with frequent/numerous ones; in particular, we illustrate in Section \ref{sec:perfcomp}
that the popular BIC \citep{y88} and mBIC \citep{zs07} penalties can perform poorly
for signals with frequent change-points. An interesting approach to data-driven
penalty selection is proposed in \cite{bm01} and \cite{bm06}, who define the so-called minimal penalty;
two approaches to estimating this quantity, referred to as `dimension jump' and `data-driven
slope estimation' are described in \cite{bmm12}. Both offer impressive performance for 
short signals with a limited number of jumps, but their performance degrades for signals
with frequent jumps (more so for the dimension jump approach). There is also computational
price to pay for selecting the penalty in a data-driven way: the `data-driven slope estimation'
approach is particularly slow for long signals.
We illustrate these points in Section 
\ref{sec:simstu}.
\cite{l05a}, motivated by the problem of selecting a suitable penalty constant from the data,
proposes a heuristic algorithm for estimating the number of change-points, which examines
the second differential of the empirical loss of the piecewise-constant model fit as a function of
the number of change-points. The approach requires the provision of the maximum number of
change-points and a threshold parameter whose value appears critical to the success of the
procedure; the theoretical properties of the method are not investigated.
We review these and some related approaches to adaptive penalty choice in Section
\ref{sec:compad} in more detail.

Another class of approaches to the multiple change-point detection problem is one
in which change-points are estimated one by one, in the order in which they appear in the data.
\cite{hs01} and \cite{km11} propose the ``moving sum'' (MOSUM) technique,
which requires the choice of an extra bandwidth parameter, but the latter requirement is 
circumvented in the {\bf mosum} R package \citep{mck18}, which produces estimates
combined over a range of automatically chosen bandwidths.
The pseudo-sequential procedure of \cite{venkatraman1993}, as well as the method of
\cite{r15} are based on an adaptation of online detection algorithms to a posteriori situations and work
by bounding the Type I error rate of falsely detecting change-points.
The `Isolate-Detect' method of \cite{af18} is also pseudo-sequential in nature, avoids to some
extent the issue of bandwidth/span selection, and offers particularly fast implementation
for signals with frequent change-points. However, from the performance point of view, such signals tend to pose major difficulties for many
methods in this category, an issue magnified by the fact that many such techniques rely critically on 
the accurate estimation of a threshold against which to judge the significance of each change-point
candidate, but the magnitude of this threshold can be difficult to estimate well 
in the presence of many change-points. We demonstrate this phenomenon in
Section \ref{sec:modsel}.

Binary segmentation is one of the most popular approaches to multiple change-point detection.
Heuristically, it works by fitting the best model with a single change-point to data $X_t$, and if the
estimated change-point is deemed to be significant, the same procedure is then repeated to the left and
to the right of the estimate. Binary segmentation is fast, conceptually simple and easy to implement, but tends not to work
well for signals with frequent change-points, the main reason being that for such signals there are no guarantees that
the single-change-point model fit at each stage of the procedure will behave well in the presence of
multiple change-points in the data segment considered.
Possibly the first work to propose binary segmentation in a stochastic process
setting is \cite{v81}, who shows consistency of binary segmentation for the number and locations
of change-points for a fixed $N$. \cite{venkatraman1993} offers a proof of the consistency of binary segmentation
for $N$ and the change-point locations, even for $N$ increasing with $T$, albeit
with sub-optimal assumptions and rates for the locations. In a setting 
similar to \cite{v81} (for a fixed $N$ and with $\varepsilon_t$ 
following a linear process), binary segmentation is considered in \cite{b97}. \cite{ccs11} provide a proof of
consistency of binary segmentation for the number of change-points in the case of a fixed $N$
and $\varepsilon_t$ i.i.d. Gaussian. Binary segmentation is used for univariate
time series segmentation in \cite{fsr11} and \cite{cf12}, and for multivariate, possibly
high-dimensional time series segmentation in \cite{cf12a}.
Circular Binary Segmentation \citep{ovlw04, vo07}, Wild Binary Segmentation \citep{f14a} 
and the Narrowest-Over-Threshold method \citep{bcf16}
are designed to improve the
performance of binary segmentation. We return to Wild Binary Segmentation later in this section.

A number of techniques do not directly fall into any of the above categories.
 \cite{w95} uses the fast
discrete wavelet transform to detect change-points. \cite{ma11} take the cumulative sum of $X_t$
and then apply the methodology described in \cite{m03} in the resulting piecewise-linear framework.
An agglomerative, bottom-up, ``tail-greedy'' method is proposed in \cite{f18}.
\cite{mj12}
present a heuristic agglomerative algorithm for multiple change-point detection as a computationally
attractive alternative to their divisive one. 
An early review of some multiple change-point detection methods (in the context
of DNA segmentation, but applicable more widely) appears in \cite{bm98}.

Wild Binary Segmentation (WBS, \citeauthor{f14a}, \citeyear{f14a}) is one of the state-of-the-art procedures for multiple
change-point detection in model (\ref{eq:model}), for signals with a small or moderate number of 
change-points. \cite{tov18} give WBS ``three stars'' (their highest
grade) for scalability to long signals, and \cite{wyr18} show certain assumption- and error-rate-optimality of WBS.
WBS is designed to deal with the shortcomings of binary segmentations
(described above) in the following manner.
At the start of the procedure, $M$ subsamples $[X_s, \ldots, X_e]$ of the data, living on intervals $[s, e]$, are drawn, with
the start- and end-points $s$ and $e$ selected randomly, independently, uniformly and with replacement.
The single-change-point model fit is then
performed on each subsample, and the first change-point candidate is selected as the one corresponding
to the best fit over the $M$ subsamples. The hope is that if $M$ is sufficiently large, then this best fit will come from a subsample
containing at most one change-point and will therefore
lead to the accurate estimation of the given change-point. If the first detected change-point is considered
significant, the same procedure is repeated to the left and to the right of it, re-using (for speed) the previously
drawn subsamples.

Notwithstanding this improvement over plain binary segmentation, we show in Section \ref{sec:perfcomp} that the performance
of WBS deteriorates dramatically for signals with frequent change-points (as does that of many other
state-of-the-art methods). The roots of its poor performance for such signals are multiple, and can be summarised
as follows.

\begin{enumerate}
\item
The recommended default number $M$ of subsamples drawn by the WBS (the R packages {\bf breakfast } \citep{breakfast} and
{\bf wbs} \citep{wbs}
recommend $M$ not exceeding 20000) can be insufficient to ensure
adequate performance of WBS for signals with frequent change-points. Indeed, good performance can only be hoped
for if $M$ is large enough for the drawn subsamples to include (with a high probability) ones with start- and end-points immediately to the left and to the
right of each change-point; this means that the required $M$ can easily be very large for signals with frequent change-points.
 Increasing the default $M$ to ensure this is not a viable solution as this would
increase the computation time beyond what is acceptable (and is not needed for signals with no or few change-points).
In the remainder of this article, we refer to this issue as the {\em lack of computational adaptivity} of the WBS,
in the sense that the procedure does not decide automatically the number $M$ 
or the locations of the subsamples drawn.
\item
A related point is that for WBS to be able to produce an entire solution path to problem (\ref{eq:model}) (i.e. estimated
models with $0, 1, \ldots, T-1$ change-points), we would have to draw all possible subsamples of the data.
This would result in a procedure of a cubic computational complexity in $T$, which is computationally inadmissible. Any value of $M$
less than this upper limit would not guarantee the computability of the entire solution path.
If a solution path algorithm in multiple change-point detection problems does not produce an entire solution path, we refer to this algorithm
as {\em incomplete}.
The use of incomplete solution path algorithms is particularly risky
in frequent change-point scenarios, as it may lead to obtaining a partial solution path that is shorter than the true number of change-points,
in which case any model selection criterion, however good, will not be able to recover all the change-points. We illustrate this important
phenomenon in Section \ref{sec:inc}.
\item
\cite{f14a} proposes two model selection criteria for the WBS: one based on thresholding, and the other based on the
use of the sSIC (a penalty close to the standard BIC/SIC). Both approaches fail in frequent change-point scenarios.
Thresholding does so because its success critically depends on the user's ability to accurately estimate the variance of the
noise $\varepsilon_t$, but this task can be difficult in frequent change-point settings, which we show in Section \ref{sec:modsel}.
BIC fails because it places too heavy a penalty on frequent-change-point solutions to the estimation problem \ref{eq:model}.
The failure of BIC in frequent change-point scenarios occurs for other change-point search algorithms too (see Section \ref{sec:perfcomp}),
not just WBS.
\end{enumerate}

The objective of this paper is to introduce the ``Wild Binary Segmentation 2" (WBS2) solution path algorithm 
for multiple change-point problems, and the ``Steepest Drop to Low Levels" (SDLL) model selection for WBS2.
The resulting multiple change-point detection method, labelled WBS2.SDLL, addresses the above shortcomings of WBS in two
separate ways.

Firstly, WBS2 produces an entire solution path, i.e. sequences of solutions to the change-point detection problem with
$0, 1, \ldots, T-1$ change-points. Among them, WBS2 guarantees (with probability approaching one with $T$) that the solution with the
true number $N$ of change-points is such that the estimated change-point locations $\tilde{\eta}_1, \ldots, \tilde{\eta}_N$ are
near-optimally close to the true change-points $\eta_1, \ldots, \eta_N$. Because it produces the entire solution path, the WBS2 is
(unlike WBS), 
by defintion, a complete procedure, and its completeness is achieved in a computationally efficient way as follows. In the first pass through
the data, WBS2 draws only a small number $\tilde{M}$ of data subsamples, and marks the first change-point candidate as the arg-max
of the resulting $\tilde{M}$ absolute CUSUMs. It then uses this change-point candidate
to split its domain of operation into two, and again recursively draws $\tilde{M}$ subsamples to the left and to the right of this change-point
candidate, and so on.
Therefore, WBS2 adaptively decides where to recursively draw the next subsamples, based on the change-point candidates detected so far
(to re-express this is the language of machine learning, WBS2
learns the locations of the subsamples to draw automatically and ``on the fly", i.e. as it proceeds through the data and discovers the signal).
This is in contrast to the standard WBS, which (non-adaptively) pre-draws all its $M$ subsamples at the start of the procedure.
We note that in practice, $\tilde{M}$ is of orders of magnitude
smaller than $M$; in the numerical sections of this paper we will be using $\tilde{M} = 100$. WBS2 draws subsamples on every current
subdomain of the data as long as its length is $\ge 2$, which ensures completeness. By contrast, WBS
(non-adaptively) pre-draws all its $M$ subsamples at the start of the procedure, which leads to its incompleteness except if
all possible subsamples are drawn, which is computationally infeasible for even moderately long signals.

Secondly, the proposed SDLL procedure for estimating $N$ from the WBS2 solution path 
does not use a penalty, and only uses thresholding as a certain secondary check,
which is in contrast to the use of thresholding as a primary model selection tool in WBS. This means that the magnitude of the threshold in WBS2.SDLL does
not need to be estimated to a high degree of accuracy (which is difficult to do in frequent change-point scenarios).
SDLL works by monitoring the ratios of the the pairs of consecutive maximum absolute CUSUM statistics
returned by the WBS2 solution path, arranged in decreasing order. The procedure then selects the largest ratio for which the
denominator falls under a pre-specified threshold. This is motivated by the fact that those absolute maximised CUSUMs that carry information about the $N$
change-points have the same order of magnitude between them, which is larger than the order of magnitude of those CUSUMs that correspond to the 
no-change-points sections of the data. The presence of this ``drop'' in the magnitude of the absolute maximised 
CUSUMs gives the SDLL criterion its name.

Our numerical experiments suggest that unlike the standard WBS, the resulting WBS2.SDLL procedure
offers good detection accuracy and fast execution times for signals with
frequent change-points. We demonstrate that in doing so, WBS2.SDLL significantly outperforms the state of the art
for this signal class. Unlike many of its competitors, WBS2.SDLL does not require the choice of a penalty or the specification of the maximum number of
change-points, and is easy to calibrate to return zero estimated change-points for constant signals with a required
probability. Both the WBS2 and the SDLL components are conceptually simple and easy to code.

The paper is organised as follows. In Section \ref{sec:motiv}, we illustrate the poor performance of the
state of the art for signals with frequent change-points, and we investigate the reasons for WBS's poor
performance in frequent change-point scenarios. Section \ref{sec:wbs2} motivates and proposes WBS2 and SDLL and investigates their
theoretical properties, as well as offering a deeper comparison of WBS2.SDLL to the existing literature. 
Our practical recommendations and a comparative simulation study are in
Section \ref{sec:num}. Section \ref{sec:ukhpi} describes the application of WBS2.SDLL to a London house
price dataset. The proofs of our theoretical results are in the Appendix.

\section{Motivation}
\label{sec:motiv}

\subsection{Frequent change-points: performance of the state of the art}
\label{sec:perfcomp}

To motivate the need for our new methodology, we start by illustrating the poor performance of WBS and other
state-of-the-art multiple change-point detection methods for signals with frequent change-points, using
the example of the \verb+extreme.teeth+ signal defined below.
We use this signal as a running example throughout the paper to motivate several of our developments,
which are then however shown to hold for a wide class of signals both in theory and in practice.
Throughout the paper, we work with methods available from the R
CRAN repository, and for each method tested, we use the default parameter settings unless stated
otherwise. For techniques that require the specification of the maximum number of change-points, we set this to the
(rounded) length of the input signal divided by three; in all our examples, the true number of change-points is (well) within
the interior of this range. Our R code enabling replication of the results from this paper is available from
\url{https://github.com/pfryz/wild-binary-segmentation-2.0}. We now list the techniques tested
in addition to our new WBS2.SDLL methodology. 

\begin{enumerate}
\item
PELT+BIC and PELT+mBIC: the PELT technique from the R package {\bf changepoint} \citep{khe16} with the BIC and mBIC penalties,
respectively. The methodology is described in \cite{kfe12}.
\item
MOSUM: the multiscale MOSUM technique with localised pruning from the R package {\bf mosum} \citep{mck18}. (We
do not give results for the multiscale MOSUM technique with bottom-up bandwidth-based merging from the same package as it
frequently returned warnings in the examples we tested.)
\item
ID: the `Isolate-Detect' technique from the R package {\bf IDetect} \citep{af18a}, described in \cite{af18}.
\item
FDRSeg: the multiscale FDR-based technique from the R package {\bf FDRSeg} \citep{ls17}, proposed in
\cite{lm16}.
\item
S3IB: the technique from the R package {\bf Segmentor3IsBack} \citep{crk16}.
It implements a fast algorithm for
minimising the least-squares cost function for change-point detection, as described in
\cite{r10}. The best model is selected via the ``oracle'' criterion of \cite{l05},
which uses the so-called slope heuristics.
\item
SMUCE: the multiscale FWER-based technique from the R package {\bf stepR} \citep{phsa18},
proposed in \cite{fms14}.
\item
CUMSEG: the linearisation-based technique from the R package {\bf cumSeg} \citep{m12},
described in \cite{ma11}.
\item
FPOP: the `functional pruning and optimal partitioning' technique from the R package {\bf fpop} \citep{rh16}
with the BIC penalty. The methodology is described in \cite{mhrf14}.
\item
DDSE and DJUMP: the ``data-driven slope estimation'' and ``dimension jump" techniques from 
the R package {\bf capushe} \citep{abbmm16}, with computation based also on the R package
{\bf jointseg} \citep{pjrn17}. The methodology is described in \cite{bmm12}.
\item
TGUH: the tail-greedy, bottom-up agglomerative technique from the R package {\bf breakfast} \citep{breakfast}.
The methodology is described in \cite{f18}.
\item
WBS-C1.0, WBS-C1.3 and WBS-BIC: the Wild Binary Segmentation methods with the model selection determined
via thresholding with constants $C=1.0$ and $C=1.3$ and the BIC (respectively), as
specified in \cite{f14a} and implemented in the R package {\bf breakfast}
\citep{breakfast}. See Section \ref{sec:wbsdef} for their definitions.
\end{enumerate}

We do not include the (non-parametric) methodology implemented in the R package {\bf ecp} \citep{jm14} as this 
procedure does not appear to be able to return zero estimated change-points. We also do not provide results for the 
R package {\bf strucchange} \citep{zlhk02} because of the slow speed of the methodology implemented therein.

\begin{figure}[h]
\begin{subfigure}{0.5\textwidth}
  \centering
  \includegraphics[width=.9\linewidth]{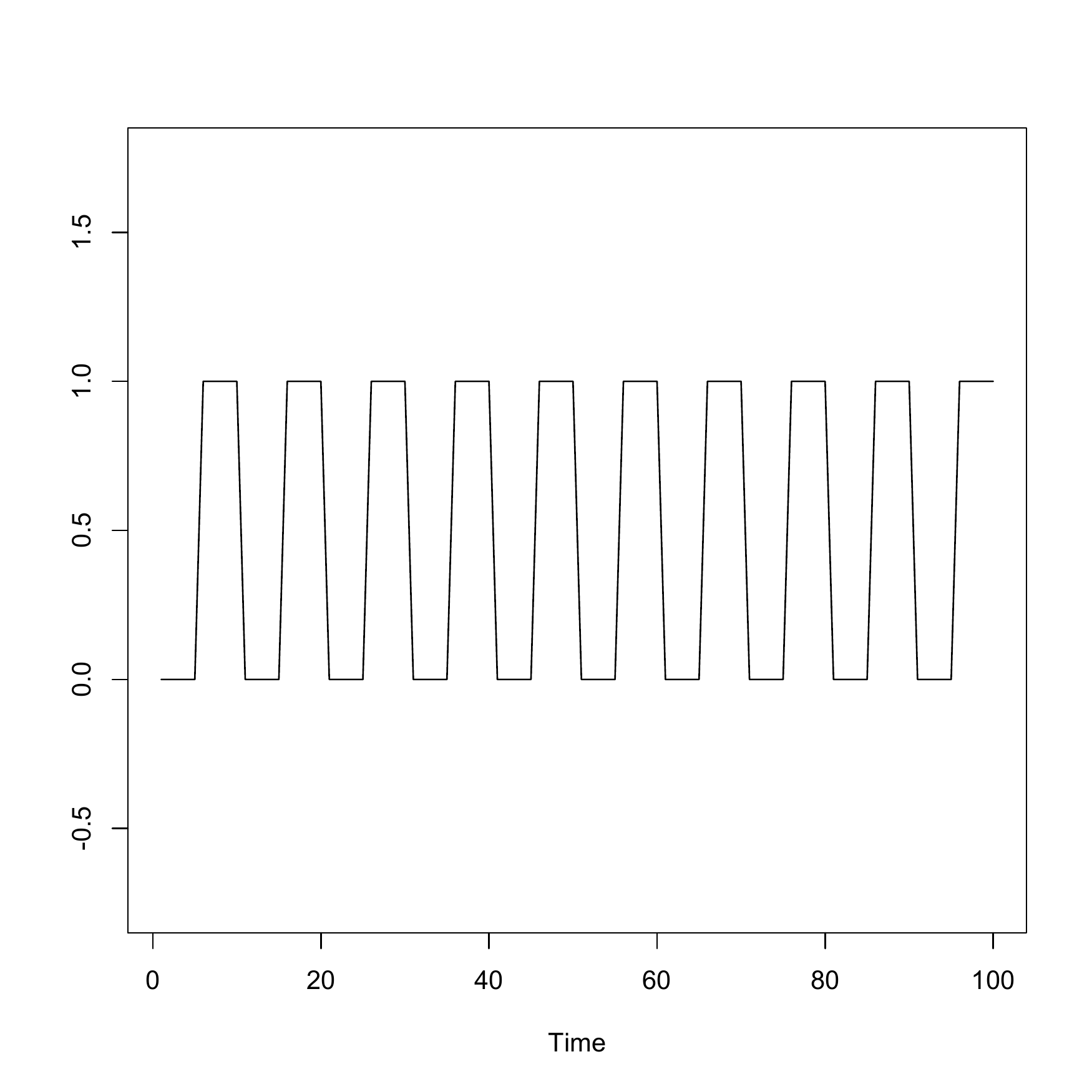}
\end{subfigure}
\begin{subfigure}{0.5\textwidth}
  \centering
  \includegraphics[width=.9\linewidth]{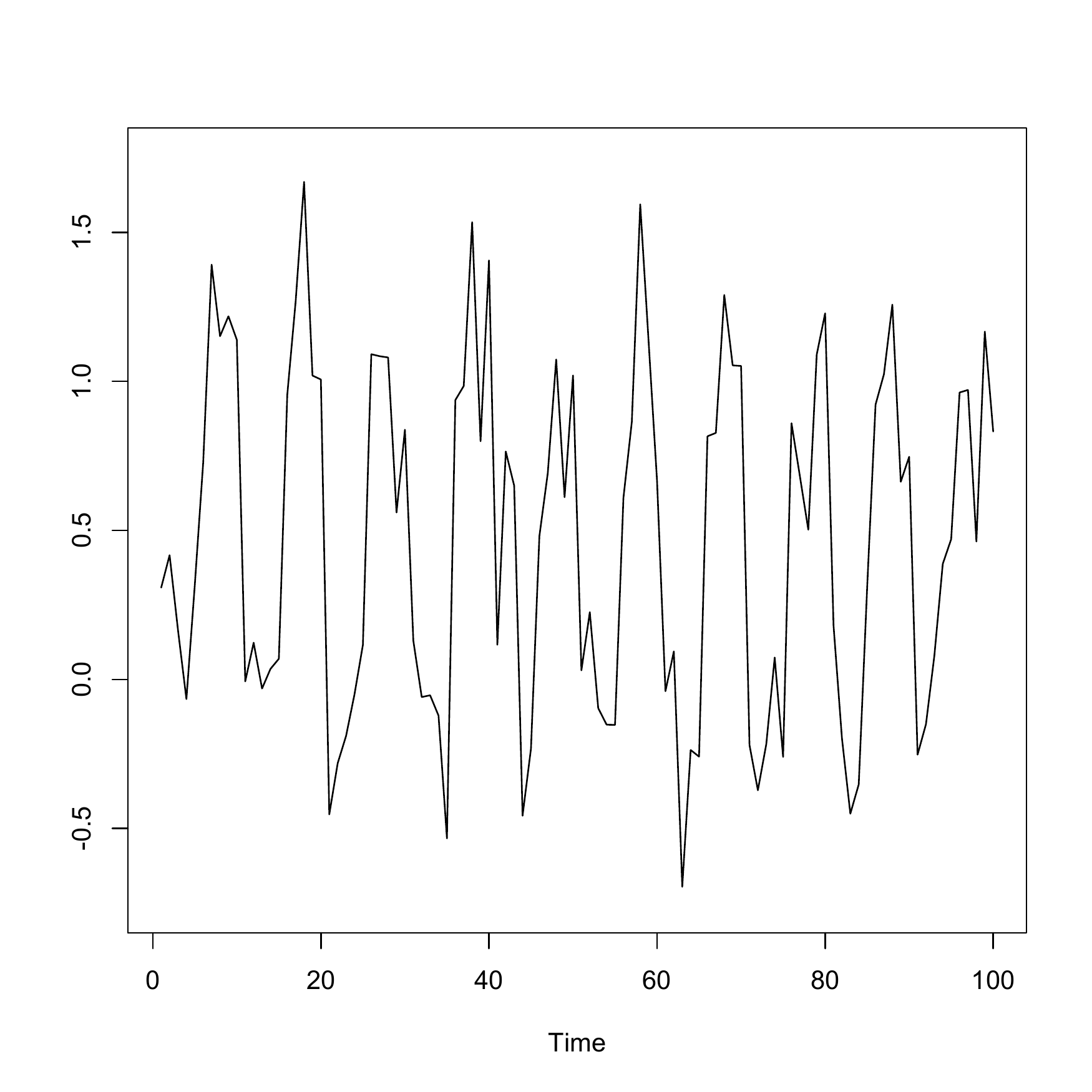}
\end{subfigure}
\caption{The first 100 observations of the {\tt extreme.teeth} signal (which continues in the same manner for $t = 1, \ldots, 1000$);
left: with no noise; right: with additive i.i.d. Gaussian noise with mean zero and $\sigma = 0.3$.}
\label{fig:ext_t}
\end{figure}

Our example in this section is the signal $f_t, t = 1, \ldots, T=1000$, referred to as \verb+extreme.teeth+ and
defined as follows: $f_t = 0$ if $1 \le (t\,\, \text{mod}\,\, 10) \le 5$; $f_t = 1$ if $(t\,\, \text{mod}\,\, 10) \in \{0, 6, 7, 8, 9\}$.
We simulate 100 independent copies of $X_t$ in model (\ref{eq:model}) with $f_t$ being the \verb+extreme.teeth+
signal and $\varepsilon_t \sim \,\,\text{i.i.d.}\,\,N(0, 0.3^2)$. Note that $N$, the true number of change-points in $f_t$,
equals 199. Figure \ref{fig:ext_t} shows the first 100 observations of the noiseless and noisy $f_t$.

\begin{figure}[h]
\begin{subfigure}{\textwidth}
  \centering
  \includegraphics[width=.8\linewidth, height = .28\textheight]{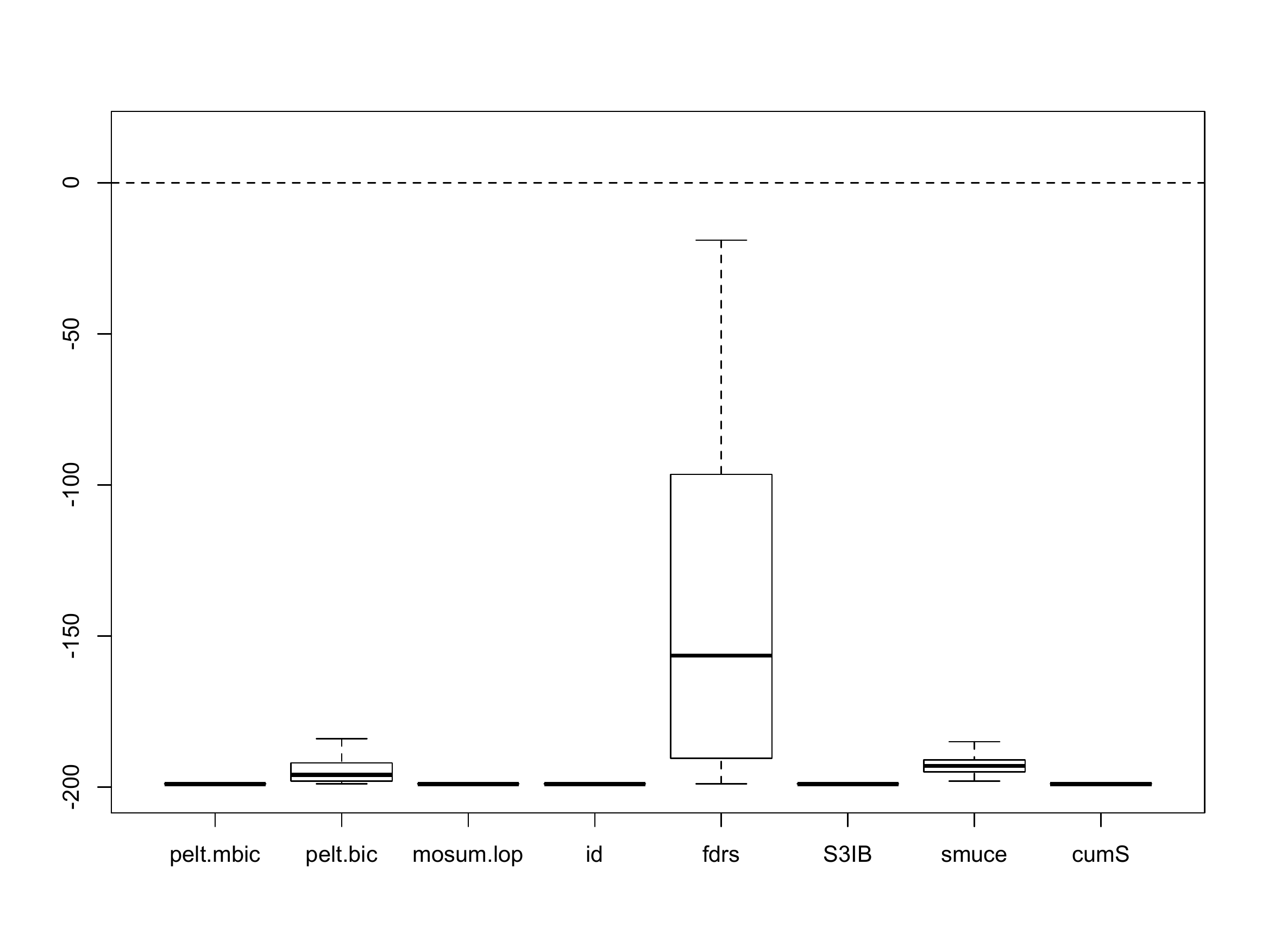}
  \label{fig:sfig3}
\end{subfigure}\\
\begin{subfigure}{\textwidth}
  \centering
  \includegraphics[width=.8\linewidth, height = .28\textheight]{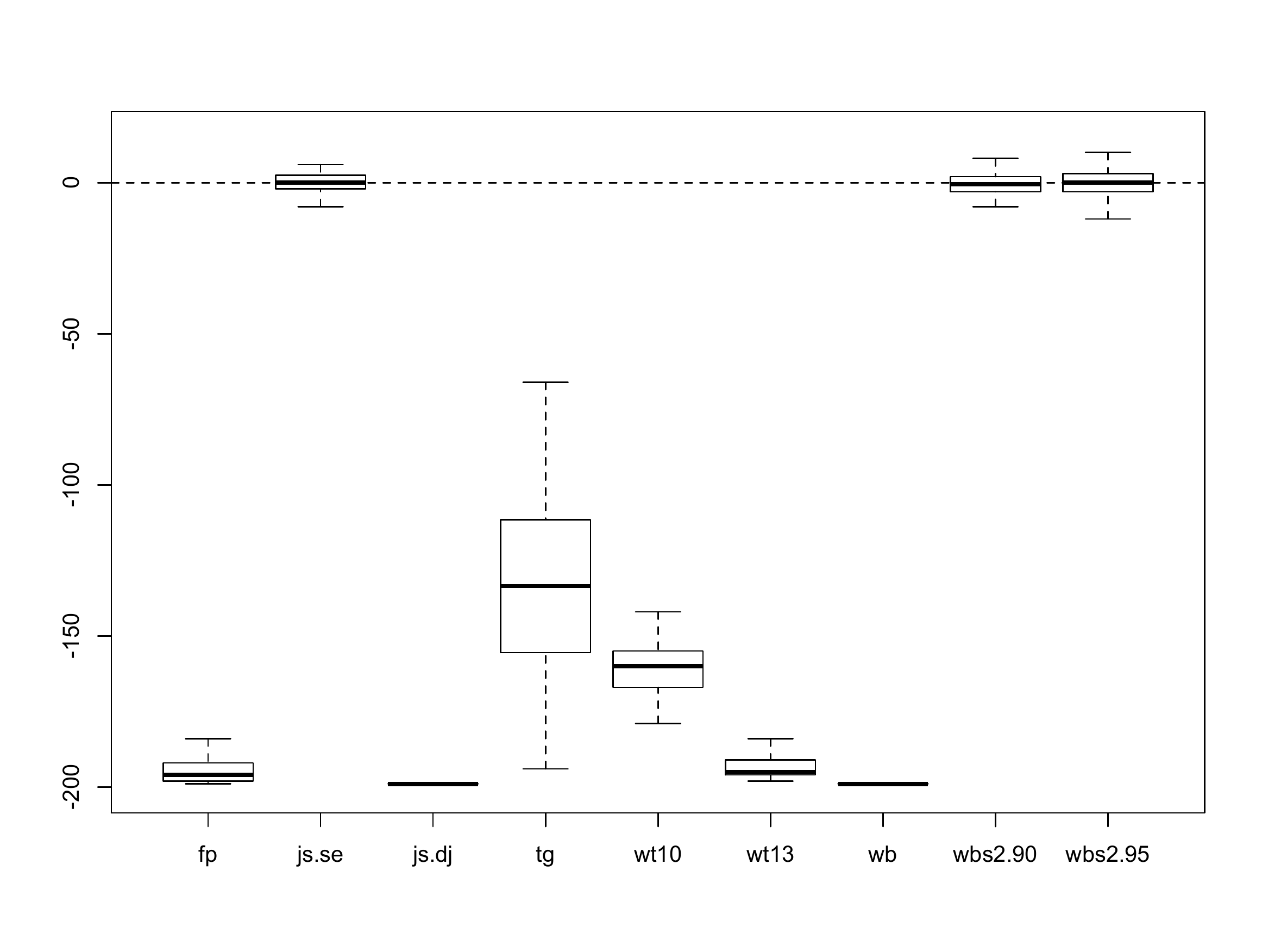}
  \label{fig:sfig4}
\end{subfigure}
\caption{From top to bottom and left to right: boxplots of $\hat{N} - N$ over 100 simulated sample paths, for the noisy {\tt extreme.teeth} signal,
for the following techniques:
PELT+mBIC, PELT+BIC, MOSUM, ID, FDRSeg, S3IB,
SMUCE, CUMSEG, FPOP,
DDSE, DJUMP, TGUH,
WBS-C1.0, WBS-C1.3, WBS-BIC
and two configurations of the WBS2 procedure proposed in this paper -- WBS2.SDLL(0.9) and WBS2.SDLL(0.95). See Section \ref{sec:perfcomp}.
}
\label{fig:ext_t_boxplots}
\end{figure}

Figure \ref{fig:ext_t_boxplots} provides evidence that most of the techniques listed above fail somewhat dramatically for this
model, with all but three being able to estimate practically none or only a small proportion of the change-points. The three best performers are
DDSE, and two versions of our method, labelled WBS2.SDLL(0.9) and WBS2.SDLL(0.95) (we provide the details later). The DDSE,
however, is over three times slower than WBS2.SDLL and estimates zero change-points for two out of the 100 sample paths,
which results in $\hat{E}_{\text{DDSE}}(\hat{N} - N)^2 \approx 800$ compared to 
$\hat{E}_{\text{WBS2.SDLL}}(\hat{N} - N)^2 \approx 20$. We return to DDSE and the other techniques in the comparative simulation
study in Section \ref{sec:simstu}. WBS-C1.0, WBS-C1.3, WBS-BIC are among the failing techniques and we now examine the reasons
for their failure.

\subsection{Wild Binary Segmentation}
\label{sec:wbsdef}

We first review WBS as proposed in \cite{f14a}.
Throughout the
paper, our main ``locator'' statistic for indicating the locations of change-point candidates is the
CUSUM statistic, defined, for the data portion 
$(X_s, \ldots, X_e)$, as 
\begin{equation}
\label{eq:ip}
\tilde{X}_{s,e}^b = \sqrt{\frac{e-b}{n(b-s+1)}}\sum_{t=s}^b X_t - \sqrt{\frac{b-s+1}{n(e-b)}}\sum_{t=b+1}^e X_t,
\end{equation}
where $s \le b < e$, with $n = e-s+1$. It is used in different ways in 
WBS and WBS2. In both these algorithms, we use
 $\arg\max_b |\tilde{X}_{s,e}^b|$ as the most likely location of a change-point in
$(f_s, \ldots f_e)$. This is the same as the location indicated by fitting the best possible piecewise-constant function
with one change-point to $(X_s, \ldots, X_e)$ via least squares. Therefore, the use of the CUSUM
statistic is equivalent to the use of maximum likelihood in model (\ref{eq:model}) in which 
$\varepsilon_t \sim$ i.i.d. $N(0, \sigma^2)$.

Denote by $F_T^M$ a set of $M$ random intervals
$[s_m, e_m]$, $m = 1, \ldots, M$, whose start- and end-points
have been drawn (independently with replacement) uniformly
from the set $\{1, \ldots, T\}$. How many intervals $M$ to draw is 
decided by the user (\cite{f14a} 
and
version 1.0.0 of the
R package {\bf breakfast} \citep{breakfast}
recommend $M=5000$ and $M = 20000$, respectively, for signal lengths $T$ not exceeding a few thousand).
An appropriate choice of $M$ is key: on the whole, the larger the value of $M$,
the more accurate but slower the procedure.
Using pseudocode, the main function of the WBS algorithm, with model selection
performed via thresholding with threshold $\zeta_T$, is defined in
the following way.

\vspace{10pt}

\begin{algorithmic}
\Function{WBS}{$s$, $e$, $\zeta_T$}
\If {$e-s < 1$}
\State STOP
\Else
	\State ${\mathcal M}_{s,e} := $ set of those indices $m$ for which $[s_m, e_m] \in F_T^M$ is such that
	$[s_m,e_m]\subseteq [s,e]$
	\State $(m_0, b_{m_0}) := \arg\max_{m\in {\mathcal M}_{s,e}, b\in\{s_m, \ldots, e_m-1\}} |\tilde{X}_{s_m,e_m}^b|$
	\If {$|\tilde{X}_{s_{m_0},e_{m_0}}^{b_{m_0}}| \ge \zeta_T$}
		\State add $b_{m_0}$ to the set of estimated change-points (with the additional side
		\State information $(m_0, s_{m_0}, e_{m_0}, |\tilde{X}_{s_{m_0}, e_{m_0}}^{b_{m_0}}|)$)
		\State \textsc{WBS}($s$, $b_{m_0}$, $\zeta_T$)
		\State \textsc{WBS}($b_{m_0}+1$, $e$, $\zeta_T$)
	\Else
		\State STOP
	\EndIf
\EndIf
\EndFunction
\end{algorithmic}

The WBS procedure (with model selection via thresholding with threshold $\zeta_T$) is launched by the call \textsc{WBS}($1$, $T$, $\zeta_T$). The WBS-C1.0
and WBS-C1.3 versions, singled out by \cite{f14a} and used in Section \ref{sec:perfcomp} above, use, respectively, $\zeta_T = \hat{\sigma} \{ 2\log\,T  \}^{1/2}$ and
$\zeta_T = 1.3\,\hat{\sigma} \{ 2\log\,T  \}^{1/2}$, where $\hat{\sigma}$ is the Median Absolute
Deviation (MAD) estimator of $\sigma = \mbox{Var}^{1/2}(\varepsilon_t)$, computed using
$\{2^{-1/2}(X_{t+1} - X_{t})\}_{t=1}^{T-1}$ on input. An alternative implementation of WBS
with thresholding first 
computes all possible change-point candidates via the call \textsc{WBS}($1$, $T$, $0$)
and then tests them, as is done in the definition of the routine \textsc{WBS}($\cdot$, $\cdot$, $\cdot$),
against the threshold $\zeta_T$.

As an alternative to thresholding, \cite{f14a} also proposes model selection via the 
``strengthened Schwarz Information Criterion" (sSIC). The sSIC penalty is close to the standard
SIC (BIC) penalty, and therefore we do not review it here in detail.
If an estimated change-point arises via the maximisation of $|\tilde{X}_{s_m,e_m}^b|$ over $b$, we refer to this
estimate
as $b_m$ in this paragraph.
The WBS-BIC version of the
procedure, used in Section \ref{sec:perfcomp}, initially computes all possible change-point candidates
via the call \textsc{WBS}($1$, $T$, $0$). 
It then sorts them in the order of decreasing magnitudes of $|\tilde{X}_{s_m,e_m}^{b_m}|$.
Let $m_1, m_2, \ldots, m_{\tilde{N}}$ be the indices (i.e. values of the indexing variable $m$) corresponding to this sorted order.
The WBS-BIC returns as the estimated change-points the set
$b_{m_1}, \ldots, b_{m_K}$ that yields the minimum of the BIC.

The key difference between the WBS and standard binary segmentation algorithms is that 
each change-point candidate is found by WBS by examining all intervals
$[s_m, e_m]$ contained within $[s, e]$ (rather than examining only $[s, e]$ itself, as in binary 
segmentation), and finding the overall arg-max
of the CUSUM statistics fitted on each such interval $[s_m, e_m]$. The reason why this improves
on the standard binary segmentation is that if $M$ is large enough, each true change-point is guaranteed (with probability
approaching one in $T$) to be contained
within at least one of the intervals $[s_m, e_m]$ which contains only that single change-point.
This alone provides suitable guarantees for the size of the largest maximum of the CUSUM
statistics, and is enough to yield substantial theoretical and practical advantages of WBS
over binary segmentation, as similar guarantees are unavailable in the latter method.
The collection of intervals $[s_m, e_m]$ is only drawn once at the start of the
WBS procedure; therefore, previously-drawn intervals are reused as the algorithm progresses.

\subsection{Issues affecting Wild Binary Segmentation}

\subsubsection{Incompleteness of WBS}
\label{sec:inc}

With the definition of the routine \textsc{WBS}($\cdot$, $\cdot$, $\cdot$) as in Section
\ref{sec:wbsdef}, the call \textsc{WBS}($1$, $T$, $0$)
returns all possible change-point candidates
in the sense that with $\zeta_T = 0$, the change-point candidates are not
tested against a threshold before they enter the output. Any
model selection procedure can then be applied to this output to identify the final set of estimated
change-points.

The call \textsc{WBS}($1$, $T$, $0$) is defined recursively. The equivalent 
non-recursive procedure for producing the same set of change-point candidates can be described as follows.
For each $[s_m, e_m] \in F_T^M$, let $b_m = \arg\max_{b\in\{s_m, \ldots, e_m-1\}} |\tilde{X}_{s_m,e_m}^b|$.
Initially let ${\mathcal C} := \{ (m, s_m, e_m, b_m, |\tilde{X}_{s_m, e_m}^{b_m}|)\,\, : \,\, m=1, \ldots, M\}$
and ${\mathcal P} := \emptyset$. While ${\mathcal C} \neq \emptyset$, repeat the following pair of
operations.
\begin{enumerate}
\item
(Identification of next change-point candidate)
\begin{eqnarray*}
m_0 & := & \text{argmax}_{m\,\,:\,\, (m, s_m, e_m, b_m, |\tilde{X}_{s_m, e_m}^{b_m}|) \in {\mathcal C}} |\tilde{X}_{s_m, e_m}^{b_m}|,\\
{\mathcal P} & := & {\mathcal P} \cup ({m_0}, s_{m_0}, e_{m_0}, b_{m_0}, |\tilde{X}_{s_{m_0}, e_{m_0}}^{b_{m_0}}|).\\
\end{eqnarray*}
\item
(Removal of all intervals containing this change-point candidate)
\[
{\mathcal C} := {\mathcal C} \setminus \{ (m, s_m, e_m, b_m, |\tilde{X}_{s_m, e_m}^{b_m}|)\,\, : \,\, s_m \le b_{m_0} < e_m\}.
\]
\end{enumerate}
At the end of this process, the set ${\mathcal P}$ contains the same set of change-point candidates as the 
output of \textsc{WBS}($1$, $T$, $0$). To see the equivalence between these two
observe that the
two recursive steps 
\textsc{WBS}($s$, $b_{m_0}$, $\zeta_T$) and \textsc{WBS}($b_{m_0}+1$, $e$, $\zeta_T$) in the definition
of the WBS routine are equivalent to: after the recording of each change-point candidate $b_{m_0}$, remove from consideration
all those intervals that contain $b_{m_0}$, and only operate on those intervals that
fall entirely within $[s, b_{m_0}]$ and $[b_{m_0}+1, e]$. In the above non-recursive definition of ${\mathcal P}$, this 
removal is explictly performed in stage 2. The removal is essential for the procedure to avoid repeated estimation
of the same change-point multiple times.

Let $\tilde{N} = |{\mathcal P}|$ (where $|\cdot|$ returns the cardinality of its argument when it is a set).
${\mathcal P}$ can be viewed as an ordered set in the following sense. Let $(m_k)_{k=1}^{\tilde{N}}$
be the order in which the 5-tuples $({m_k}, s_{m_k}, e_{m_k}, b_{m_k}, |\tilde{X}_{s_{m_k}, e_{m_k}}^{b_{m_k}}|)$
enter ${\mathcal P}$. The sequence $(|\tilde{X}_{s_{m_k}, e_{m_k}}^{b_{m_k}}|)_{k=1}^{\tilde{N}}$ is non-increasing,
as its each next element is a maximum over the progressively shrinking set ${\mathcal C}$. In the remainder
of this paper, we treat ${\mathcal P}$ as an ordered set with this particular ordering.
Therefore, the ordered set ${\mathcal P}$ can be viewed as a WBS ``solution path":
the change-point candidates contained in ${\mathcal P}$ are arranged in order of decreasing importance
(i.e. decreasing magnitude of $|\tilde{X}_{s_{m_k}, e_{m_k}}^{b_{m_k}}|$),
and therefore any reasonable WBS model selection procedure would typically
choose a certain number of the first components of ${\mathcal P}$
as the final set of estimated change-points.

As a result of the repeated removal of intervals from the set ${\mathcal C}$, the cardinality $\tilde{N}$ of the solution
path ${\mathcal P}$ is typically much smaller than the cardinality $M$ of the initial set ${\mathcal C}$. This is
illustrated in the following example.

{\bf Example 2.1} \hspace{2pt} Consider a sample path simulated from the model $X_t = f_t + \varepsilon_t$, $t = 1, \ldots, T=1000$, where
$f_t$ is the \verb+extreme.teeth+ signal defined in Section \ref{sec:perfcomp} and $\varepsilon_t \sim N(0, 0.3^2)$ and
has been simulated in R on setting the random seed to 1. We simulate $M = 5000$ (a value recommended
in \cite{f14a}) intervals $[s_m, e_m]$. On completion of the computation of ${\mathcal P}$
as described in this section, we obtain $\tilde{N} = 119$. As the length $\tilde{N}$ of the solution path ${\mathcal P}$
is less than 199, the true number of change-points in $f_t$, no model selection procedure is able to recover 
all the change-points in this example: indeed, even if we were to accept all change-point candidates in ${\mathcal P}$,
we would only estimate 119 change-points. Therefore, WBS as proposed in \cite{f14a} fails in this case. (End of example.)

The above example illustrates that too small a value of $M$ can prevent WBS from working well for signals with frequent
change-points, regardless even of the model selection procedure used. Increasing the default value of $M$ in the hope of
obtaining a longer solution path ${\mathcal P}$ is not a viable option as it can easily slow down the WBS procedure beyond what is
acceptable (in particular, drawing all possible sub-intervals of $[1, \ldots, T]$ leads to WBS having computational speed
$O(T^3)$) and is unnecessary for signals with no or few change-points, which require far fewer intervals. In other words,
WBS does not adapt to the data in terms of the number $M$ or the locations $[s_m, e_m]$ of the intervals it
draws. From the user perspective, this lack of adaptivity creates two related major practical issues: 
(a) in the absence of prior information about the number and frequency of change-points,
it is unclear whether any particular value of $M$ is large enough in practice, and 
(b) setting $M$ to be very large to be ``on the safe side'' is not an option from the computational point of view.

We formalise the issue of the length of solution paths in the language of ``completeness'', which we now introduce.

\begin{defin}
\label{def:complete}
Let $A$ be a solution path algorithm in a multiple change-point detection problem which produces 
a sequence of candidate models with $0, 1, \ldots, \tilde{N}$ estimated change-points.
We refer to $A$ as {\em complete} if $\tilde{N} = T-1$, where $T$ is the length of the input data,
and {\em incomplete} otherwise.
\end{defin}

Note that $\tilde{N} = T-1$ means that every time point is a change-point candidate, so that the candidate model with
$\tilde{N} = T-1$ is maximal. The definition of completeness leads to the following proposition for WBS. The proof is straightforward, so we omit it.

\begin{prop}
The solution path ${\mathcal P}$ obtained via WBS as described in this section is guaranteed to be
complete in the sense of Definition \ref{def:complete} if and only if $\{ [s_m, e_m] \}_{m=1}^M$ represent all
possible sub-intervals of $[1, \ldots, T]$, in which case $M = T(T-1)/2$. Furthermore, if $M = O(T^2)$, then
the WBS solution path procedure is of computational order $O(T^3)$.
\end{prop}

We argue that completeness is a desirable feature of the solution path component of 
any change-point detection procedure if it is to be applied
in frequent change-point scenarios, as the presence of completeness guarantees that we avoid situations (such as the
one illustrated in this section) in which the solution path is too short to enable estimation
of all the change-points.
Completeness embodies the idea that no change-point candidate should be dismissed as inadmissible
at the ``solution path'' stage: this should only happen at the model selection stage.
For both scalability and good performance in frequent change-point scenarios, it is important for a solution path
procedure to be both fast and complete, and we have shown in this section that
WBS does not simultaneously possess these two features.
\comment{
By contrast, the WBS2 procedure proposed in this article will be both fast and complete.
}

\subsubsection{Shortcomings of thresholding and BIC as model selection for WBS}
\label{sec:modsel}

\cite{f14a} proposes two model selection criteria for WBS: one based on thresholding, and the other based on
the sSIC, an information criterion for all practical purposes equivalent to the standard SIC (a.k.a. BIC). 
In this section, we argue that both can easily be inadequate in frequent
change-point scenarios (independently of the issue of completeness described in Section \ref{sec:inc}).

The success of thresholding as a model selection criterion for WBS critically relies on the user's ability to accurately
estimate $\sigma = \mbox{Var}^{1/2}(\varepsilon_t)$. The following example illustrates that this task can be difficult
to achieve in frequent change-point models.

\begin{figure}[h]
  \centering
  \includegraphics[width=.8\linewidth, height=.4\textheight]{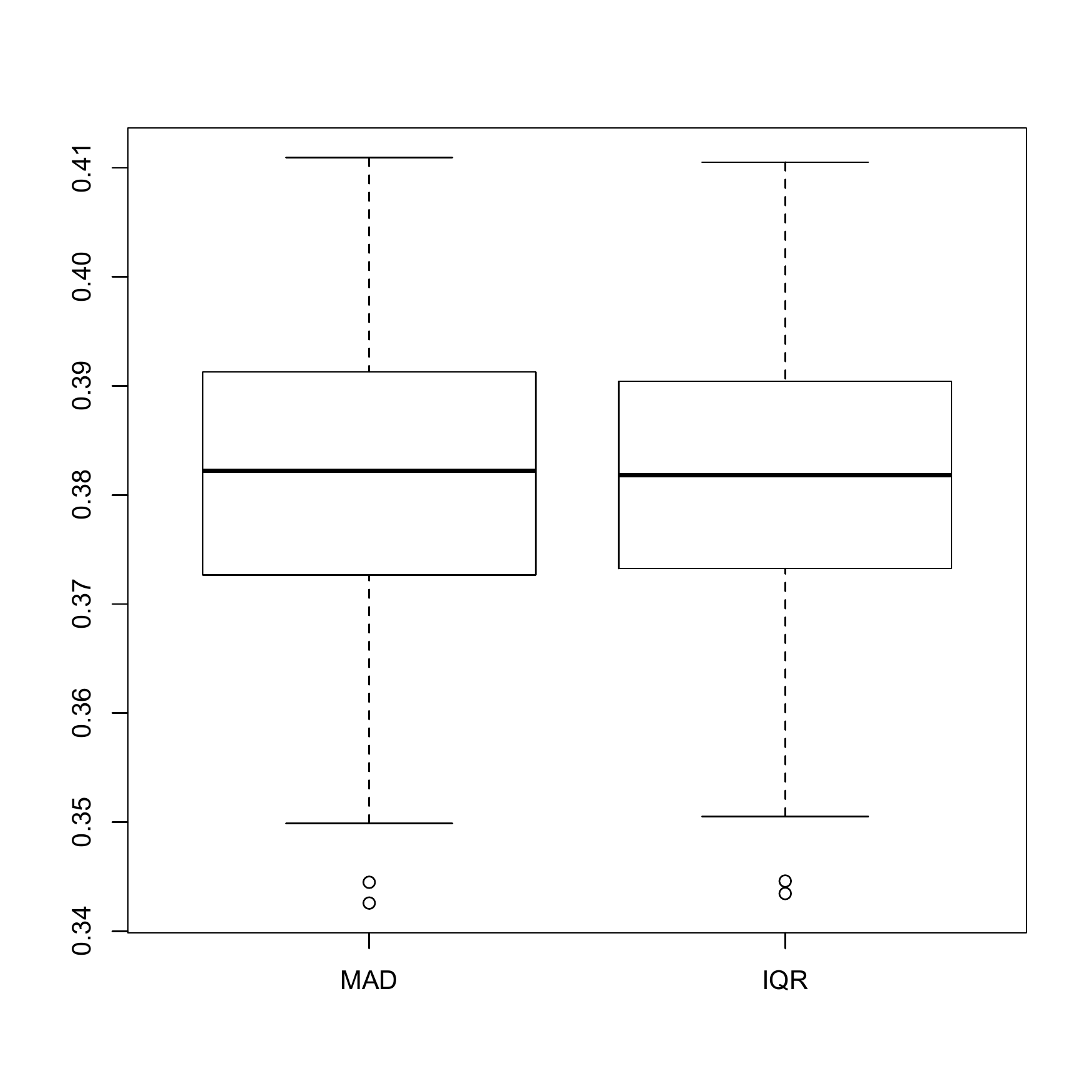}
\caption{Boxplots of the empirical distribution of the MAD (left) and IQR (right) estimators of $\sigma$ in the {\tt extreme.teeth}
model described in Section \ref{sec:modsel}, over 100 simulated sample paths. The true value of $\sigma$ is 0.3.}
\label{fig:madiqr}
\end{figure}

{\bf Example 2.2} \hspace{2pt} Consider 100 sample paths simulated from the model $X_t = f_t + \varepsilon_t$, $t = 1, \ldots, T=1000$, where
$f_t$ is the \verb+extreme.teeth+ signal defined in Section \ref{sec:perfcomp}. and $\varepsilon_t \sim N(0, 0.3^2)$.
For each simulated sample path, we evaluate the Median Absolute Deviation (MAD) and Inter-Quartile Range (IQR)
estimators (two commonly used robust estimators) of $\sigma=0.3$ calibrated for the Gaussian distribution, computed with $\{2^{-1/2}(X_{t+1} - X_{t})\}_{t=1}^{T-1}$
on input. Figure \ref{fig:madiqr} shows that both estimators are heavily biased upwards on this example and overestimate
$\sigma$ by around 25\%. The degree of this overestimation prevents thresholding from being an effective model selection tool in this model.
(End of example.)

In addition, there is ample empirical evidence that the BIC is a poor model selector in frequent change-point scenarios, also
for search algorithms other than WBS. For instance, we showed in Section \ref{sec:perfcomp} that the PELT method \citep{kfe12}
equipped with the BIC stopping rule failed on the \verb+extreme.teeth+ example. In brief, this is due to the fact that the
BIC term places too heavy a penalty on frequent-change-point solutions.

It is tempting to ask at this point whether choosing a penalty adaptively from the data (so that possibly different penalties are
applied in infrequent and frequent change-point scenarios) may provide a successful remedy to this issue. Based on our empirical
experience, the best such approach that we have come across is the ``data-driven slope estimation'' technique implemented in the
R package {\bf capushe} \citep{abbmm16} and described in \cite{bmm12}. However, we show in Section \ref{sec:simstu} that
this approach scales poorly to long signals and is outperformed by WBS2.SDLL in the frequent change-point examples
we tested.

\comment{

This article proposes a new model selection approach, termed the ``Steepest Drop to Low Levels'' to the multiple 
change-point detection problem. It is presented in the context of,
and integrated into, the WBS2 procedure, but our expectation is that it should be applicable more widely.
It avoids the issues mentioned in this section by (a) not using a penalty and (b) using thresholding only as a certain
secondary-check device, which relieves it from having to use a very accurate estimate of $\sigma$.

}

\section{Wild Binary Segmentation 2 and Steepest Drop to Low Levels}
\label{sec:wbs2}

\subsection{WBS2: data-adaptive computation of complete solution path}

\subsubsection{Computation of the WBS2 solution path}
\label{sec:solpath}

This section describes how the WBS2 solution path is computed. This is the first of the two ingredients
of the WBS2.SDLL procedure; the other is the application of the new ``Steepest Drop to Low Levels" model selection
criterion, described in Section \ref{sec:sdll}.

We start with a heuristic description, followed by a formal algorithmic definition. In the first pass through
the data, WBS2 draws a small number $\tilde{M}$ of data subsamples, with start- and end-points chosen
randomly, uniformly and independently with replacement from the set $\{1, \ldots, T\}$. WBS2 marks the first change-point candidate as the arg-max
of the resulting $\tilde{M}$ absolute CUSUMs. It then uses this change-point candidate
to split its domain of operation $\{1, \ldots, T\}$ into two, and again recursively draws $\tilde{M}$ subsamples to the left and to the right of this change-point
candidate, and so on. Eventually, on short sub-domains, the number of all possible subsamples will be less than $\tilde{M}$,
in which case WBS2 will draw all possible subsamples of each such sub-domain. WBS2 stops and takes no action on a given current
sub-domain only if its length is 1. Having completed this step, WBS2 then sorts all thus-obtained change-point candidates in the
decreasing order of their corresponding absolute CUSUMs. The resulting object is referred to as the WBS2 solution path.
Algorithmically, the main ingredient of the computation of the WBS2 solution path is the routine \textsc{WBS2.Sol.Path}, defined
recursively as follows.

\begin{algorithmic}
\Function{WBS2.Sol.Path}{$s$, $e$, $\tilde{M}$}
\If {$e-s < 1$}
\State STOP
\EndIf
\If {$\tilde{M} \ge \frac{1}{2}(e-s+1)(e-s)$}
\State $\tilde{M} := \frac{1}{2}(e-s+1)(e-s)$ 
\State Draw all intervals $[s_m, e_m] \subseteq [s, s+1, \ldots, e]$, $m=1, \ldots, \tilde{M}$, s.t. $e_m-s_m > 1$.
\Else
\State Randomly draw intervals $[s_m, e_m] \subseteq [s, s+1, \ldots, e]$, $m=1, \ldots, \tilde{M}$, s.t. $e_m-s_m > 1$
\State and $s_m, e_m$ drawn uniformly and independently with replacement.
\EndIf
\State $(m_0, b_{m_0}) := \arg\max_{m = 1, \ldots, \tilde{M},\, b\in\{s_m, \ldots, e_m-1\}} |\tilde{X}_{s_m,e_m}^b|$
\State Add $(s_{m_0}, e_{m_0}, b_{m_0}, |\tilde{X}_{s_{m_0}, e_{m_0}}^{b_{m_0}}|)$ to the (unsorted) solution path $\tilde{{\mathcal P}}$.
\State \textsc{WBS2.Sol.Path}($s$, $b_{m_0}$, $\tilde{M}$)
\State \textsc{WBS2.Sol.Path}($b_{m_0}+1$, $e$, $\tilde{M}$)
\EndFunction
\end{algorithmic}

The WBS2 solution path $\tilde{{\mathcal P}}$ is computed in two stages, defined below.

\begin{enumerate}
\item (Identification of change-point candidates)

Initialise $\tilde{{\mathcal P}} := \emptyset$.

Add elements to the (unsorted) WBS2 solution path $\tilde{{\mathcal P}}$ by executing the command \textsc{WBS2.Sol.Path}($1$, $T$, $\tilde{M}$). We state
in Section \ref{sec:spth} that the length of the thus-obtained sequence $\tilde{\mathcal P}$ is $T-1$.

\item (Sorting of the WBS2 solution path)

Rearrange the elements of $\tilde{{\mathcal P}}$ so that the sequence of its 4th components, denoted by $(|\tilde{X}_{s_{k}, e_{k}}^{b_{k}}|)_{k=1}^{T-1}$,
is non-increasing, breaking any ties at random. 
Denote the elements of the resulting (sorted) sequence $\tilde{{\mathcal P}}$ by $(s_{k}, e_{k}, b_{k}, |\tilde{X}_{s_{k}, e_{k}}^{b_{k}}|)_{k=1}^{T-1}$.

\end{enumerate}

The philosophy of WBS2 is fundamentally different from WBS. In WBS, all $M$ intervals are pre-drawn before the launch of the procedure, and there
needs to be a sufficient number of them for {\em all} change-points to be detectable from this single draw. By contrast, in WBS2, with each draw of $\tilde{M}$
intervals, we only hope to detect {\em one} of the remaining undetected change-points (if there are any). For this reason, 
$\tilde{M}$ will typically be much smaller than $M$. We comment in Section \ref{sec:spth} on the theoretical orders of magnitude
of $\tilde{M}$; for now suffice it to say that we use $\tilde{M} = 100$ in all numerical examples of this paper.

In comparison to WBS, the computation of the WBS2 solution path is data-adaptive. WBS2 does not use the data in drawing its $M$ intervals; by contrast,
in WBS2, the location of the domain of each next batch of $\tilde{M}$ intervals is determined by the locations of the previously detected change-point candidates.
One consequence of this computational adaptivity of WBS2 is that on average, the intervals drawn by WBS2 rapidly become shorter as the procedure progresses,
as the procedure operates on shorter and shorter sub-domains of $[1, \ldots, T]$. This makes WBS2 fast; we comment on the theoretical computational complexity
of the WBS2 solution path in Section \ref{sec:spth} and actual computation times are given in Section \ref{sec:simstu}.

Finally, we note that the computation of the WBS2 solution path requires no other parameters besides $\tilde{M}$, and that it is natural for it to be defined
(and indeed implemented) recursively.

\subsubsection{Theoretical properties of the WBS2 solution path}
\label{sec:spth}

We introduce the following technical assumption.

\begin{assumption}
\label{ass:tech}
\hspace{0pt}
\begin{enumerate}
\item[(a)]
The sequence $(\varepsilon_t)_{t=1}^T$ is i.i.d. $N(0, \sigma^2)$.
\item[(b)]
With the additional notation $\eta_0 = 1$, $\eta_{N+1} = T+1$, the minimum spacing between change-point
satisfies $\min_{i=1, \ldots, N+1} (\eta_{i} - \eta_{i-1}) \ge \delta_T$, where $\delta_T = \delta\,T$ with $\delta > 0$.
\item[(c)]
$|f_t| \le \bar{f} < \infty$ for $t=1, \ldots, T$.
\item[(d)]
The magnitudes $f'_i = |f_{\eta_i} - f_{\eta_i - 1}|$ of the jumps satisfy $\min_{i=1,\ldots,N} f'_i \ge \underline{f}_T \ge C T^{\kappa}$ for
some $C > 0$ and $\kappa > -1/2$.
\end{enumerate}
\end{assumption}

Assumptions \ref{ass:tech}(a) or (b) are not strictly necessary for the procedure to work, and both have been made for technical
simplicity and to simplify the exposition as much as possible, while still enabling the demonstration of the full mechanics of the 
procedure; various extensions are in principle possible. In frequent change-point scenarios, the reader is invited to think of situations
in which $\delta$, the smallest possible spacing between neighbouring change-points as a fraction of the sample size, is small. We note, also,
that the minimum jump size in Assumption \ref{ass:tech}(d) satisfies the condition spelled out in Assumption 3.3 of \cite{f14a}.
Finally, we emphasise that it is permitted for the jump sizes $f'_i$ to have different orders of magnitude in $T$, as long as their minimum
satisfies the lower bound in Assumption \ref{ass:tech}(d).

To facilitate the statement of the theorem below, we introduce the concept of `scale' at which the WBS2 procedure operates as follows. Initially, the procedure operates on the
domain $[1, \ldots, T]$, and we refer to this stage as scale 1. It then subdivides this domain into two and operates on the two
sub-domains; this is scale 2. The scale parameter then increases by one with each subdivision.

We have the following result.

\begin{theorem}
\label{th:solpath}
Let $X_t$ follow model (\ref{eq:model}) and let Assumption \ref{ass:tech} hold. Let $N$ and $\eta_1, \ldots, \eta_N$ denote,
respectively, the number and the locations of the change-points. Let $\tilde{\mathcal P}$ be the solution path of the WBS2 
algorithm, defined in Section \ref{sec:solpath}. The following statements are true.
\begin{enumerate}
\item[(i)] 
The length of $\tilde{\mathcal P}$ is $T-1$.
\item[(ii)]
Let $\Gamma_j$ be the length of the longest sub-domain on which the WBS2 solution path algorithm operates at scale $j = 1, 2, \ldots$. Let
$J = \min \{j\, : \, \Gamma_j = 1\} - 1$. The computational complexity of the WBS2 solution path algorithm is $O(\tilde{M} J T)$.
\item[(iii)]
Let $\{  b_k \}_{k=1}^{T-1}$ be the 3rd components of the elements of $\tilde{\mathcal P}$. Sort $\{  b_k \}_{k=1}^N$ in increasing order, and
denote the thus-ordered sequence by $(\tilde{\eta}_i)_{i=1}^N$. Define the event
\begin{eqnarray*}
{\mathcal A}_T & = & \{  \max_{i=1,\ldots,N} |\eta_i - \tilde{\eta}_i|  \le C_1(\Delta) (\underline{f}_T)^{-2}\,\log\,T \quad\cap\quad  
|\tilde{X}_{s_k,e_k}^{b_k}| \gtrsim \underline{f}_T T^{1/2}\,\,\,\text{for}\,\,\, k = 1, \ldots, N\\
& \cap &  |\tilde{X}_{s_k,e_k}^{b_k}| \le C_2(\Delta) \log^{1/2} T\,\,\, \text{for}\,\,\, k = N+1, \ldots, T-1\},
\end{eqnarray*}
where $C_1(\Delta), C_2(\Delta)$ are positive constants that only depend on the positive $\Delta$ parameter below,
and the $\gtrsim$ symbol means ``of the order of or larger''. For $T$ large enough, we have
\begin{equation}
\label{eq:lowbnd}
P({\mathcal A}_T) \ge 1 - \alpha_T - \frac{1}{2}N(N+1) (2C_1(\Delta)(\underline{f}_T)^{-2}\log\,T  + 1)^2 (1 - \delta^2/9)^{\tilde{M}},
\end{equation}
where
\[
\alpha_T = 1 - P({\mathcal B}_{\Delta, T})
\]
with
\[
{\mathcal B}_{\Delta, T} = \left\{  \forall\,\, 1 \le l \le m \le T\qquad |m-l+1|^{-1/2} \left| \sum_{i=l}^m \varepsilon_i  \right| \le \sigma \{2(1+\Delta)\log\,T\}^{1/2}   \right\}.
\]
\end{enumerate}
\end{theorem}

We now comment on several aspects of Theorem \ref{th:solpath}. Part (i) means that the WBS2 solution path
procedure is complete in the sense of Definition \ref{def:complete}. In part (ii), we investigate the magnitudes of
$\tilde{M}$ and $J$ to gain a better idea of the computational complexity of the WBS2 solution path procedure.
Beginning with $\tilde{M}$, we 
now discuss the minimum number of draws $\tilde{M}$ required to ensure that $P({\mathcal A}_T)$ in part (iii)
 converges to 1 suitably fast. To match the equivalent rate for WBS (see Theorem 3.2 in \cite{f14a}), we require
\[
\frac{1}{2}N(N+1) (2C_1(\Delta)(\underline{f}_T)^{-2}\log\,T + 1)^2 (1 - \delta^2/9)^{\tilde{M}} \le T^{-1},
\]
which is equivalent to
\[
\tilde{M} \ge \log^{-1}\{(1 - \delta^2/9)^{-1}\} (\log\,T + \log\{N(N+1)/2\} + 2\log(2C_1(\Delta)(\underline{f}_T)^{-2}\log\,T+1)),
\]
and therefore $\tilde{M} = O(\log\,T)$. The maximum magnitude of $J$ depends on the location of the splits
$b_{m_0}$ within each each interval $[s,e]$ on which the routine \textsc{WBS2.Sol.Path}$(s,e,\tilde{M})$
is executed. In particular, we have the following result (the proof is straightforward, so we omit it).

\begin{prop}
\label{prop:J}
If, in each execution of the routine \textsc{WBS2.Sol.Path}$(s,e,\tilde{M})$, the split location $b_{m_0}$ is
such that 
\begin{equation}
\label{eq:balance}
\gamma \le \frac{b_{m_0} - s + 1}{e - s + 1} \le 1-\gamma,
\end{equation}
for a certain fixed $\gamma \in (0, 1)$, then $J = O(\log\,T)$.
\end{prop}
It is immediate that condition (\ref{eq:balance}) is automatically satisfied if $e-s \le 1/\gamma - 1$, i.e. for short
intervals. From the proof of Theorem \ref{th:solpath}, it is also automatically satisfied (for $\gamma$ small enough,
with probability tending to 1 with $T$) for those intervals $[s,e]$ which contain previously undetected change-points.
Since the locations of the splits $b_{m_0}$ in those intervals $[s,e]$ that do not 
contain previously undetected change-points is immaterial for the performance of the procedure, 
this in turn implies that an alternative WBS2 solution path procedure, in which the existing selection
of $b_{m_0}$ were replaced by constrained maximisation of the form
\[
(m_0, b_{m_0}) := \arg\max_{m = 1, \ldots, \tilde{M},\, b\in\{s_m, \ldots, e_m-1\}} |\tilde{X}_{s_m,e_m}^b|
\quad\text{s.t. condition (\ref{eq:balance})}
\]
would enjoy exactly the same properties as the original WBS2 solution
path procedure, plus would come with a guarantee that $J = O(\log\,T)$. However, we refrain from discussing such
a modified procedure further as it would require the choice of $\gamma$, an additional parameter.

Proposition \ref{prop:J} and the above discussion regarding the magnitude of $\tilde{M}$ imply the following corollary.
\begin{corollary}
Under the assumptions of Theorem \ref{th:solpath} and Proposition \ref{prop:J},
the computational complexity of the WBS2 solution path algorithm is $O(T\log^2 T)$.
\end{corollary}

Regarding part (iii) of Theorem \ref{th:solpath}, 
it is the statement of Lemma 1
in \cite{y88} that $P({\mathcal B}_{\Delta, T}) \to 1$ for all $\Delta > 0$, at a rate that depends on the magnitude
of $\Delta$. The uniform bound on the magnitudes of the errors $|\eta_i - \tilde{\eta}_i|$ is the same as
in WBS and is `near-optimal' in the sense described in \cite{f14a}.

We emphasise that part (iii) does not yet constitute a complete consistency result for the WBS2.SDLL procedure, as it 
does not specify a model selection process; it merely says that the model with the true number $N$ of estimated
change-points is such that the locations of all $N$ change-points are estimated near-optimally. In Section \ref{sec:sdll},
we will describe how to estimate $N$ consistently via SDLL and this will lead to
a complete consistency result (for the number and locations of estimated change-points) for the WBS2.SDLL method.

\subsection{``Steepest Drop to Low Levels" model selection}
\label{sec:sdll}

\subsubsection{Motivating example and methodology of SDLL model selection}
\label{sec:sdllmet}

This section introduces our new model selection criterion, labelled ``Steepest Drop to Low Levels'', which will be used to select a model 
along the
WBS2 solution path $\tilde{\mathcal P}$. Given $\tilde{\mathcal P}$, the model selection task is equivalent to indicating an estimate $\hat{N}$
of $N$ (the number of change-points), as the estimated change-point locations will then be given by $\{b_k\}_{k=1}^{\hat{N}}$, if $\hat{N} > 0$.
This section will define our estimator $\hat{N}$.
We start with a motivating example.

\begin{figure}[h]
\begin{subfigure}{\textwidth}
  \centering
  \includegraphics[width=.8\linewidth, height = .28\textheight]{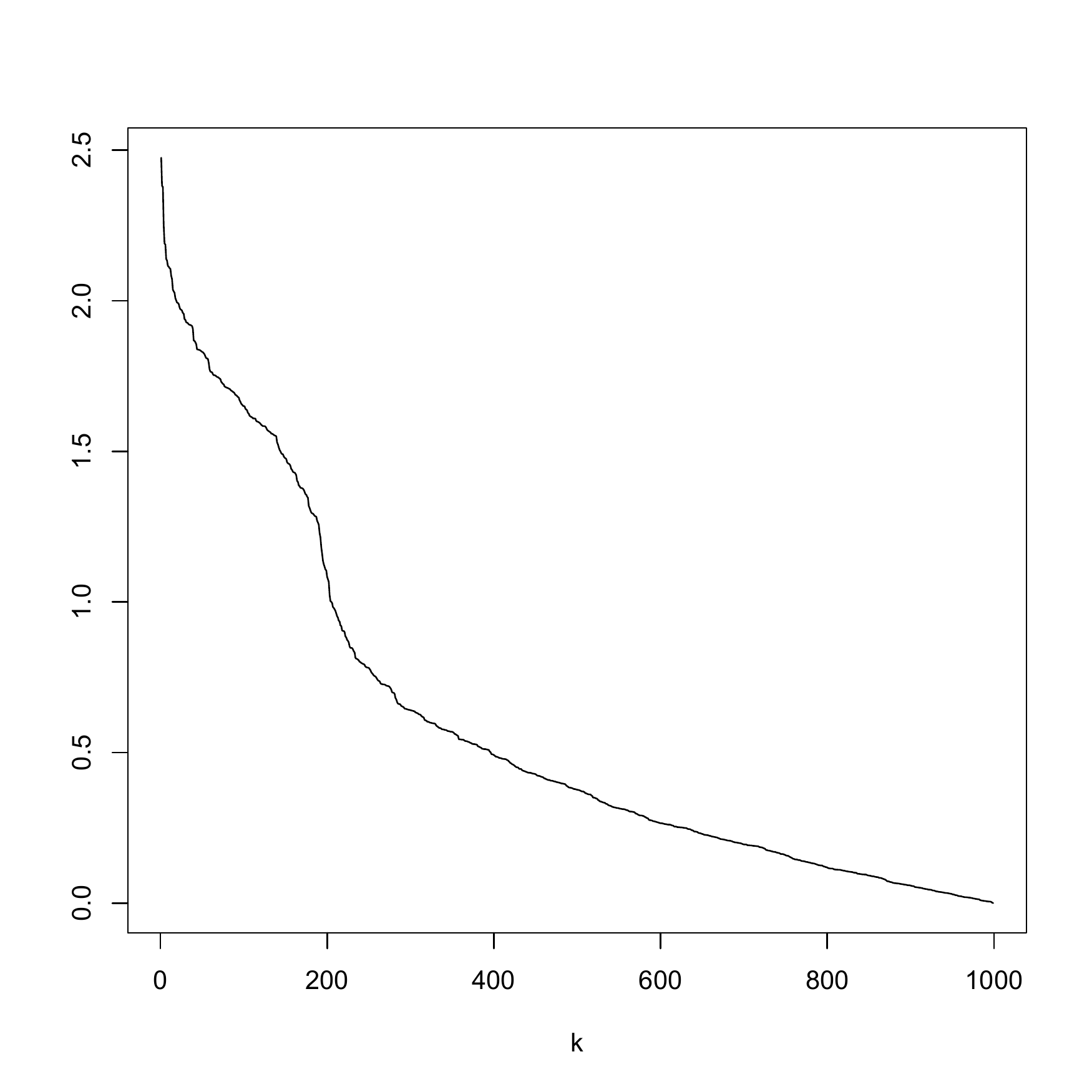}
  \label{fig:spd}
\end{subfigure}\\
\begin{subfigure}{\textwidth}
  \centering
  \includegraphics[width=.8\linewidth, height = .28\textheight]{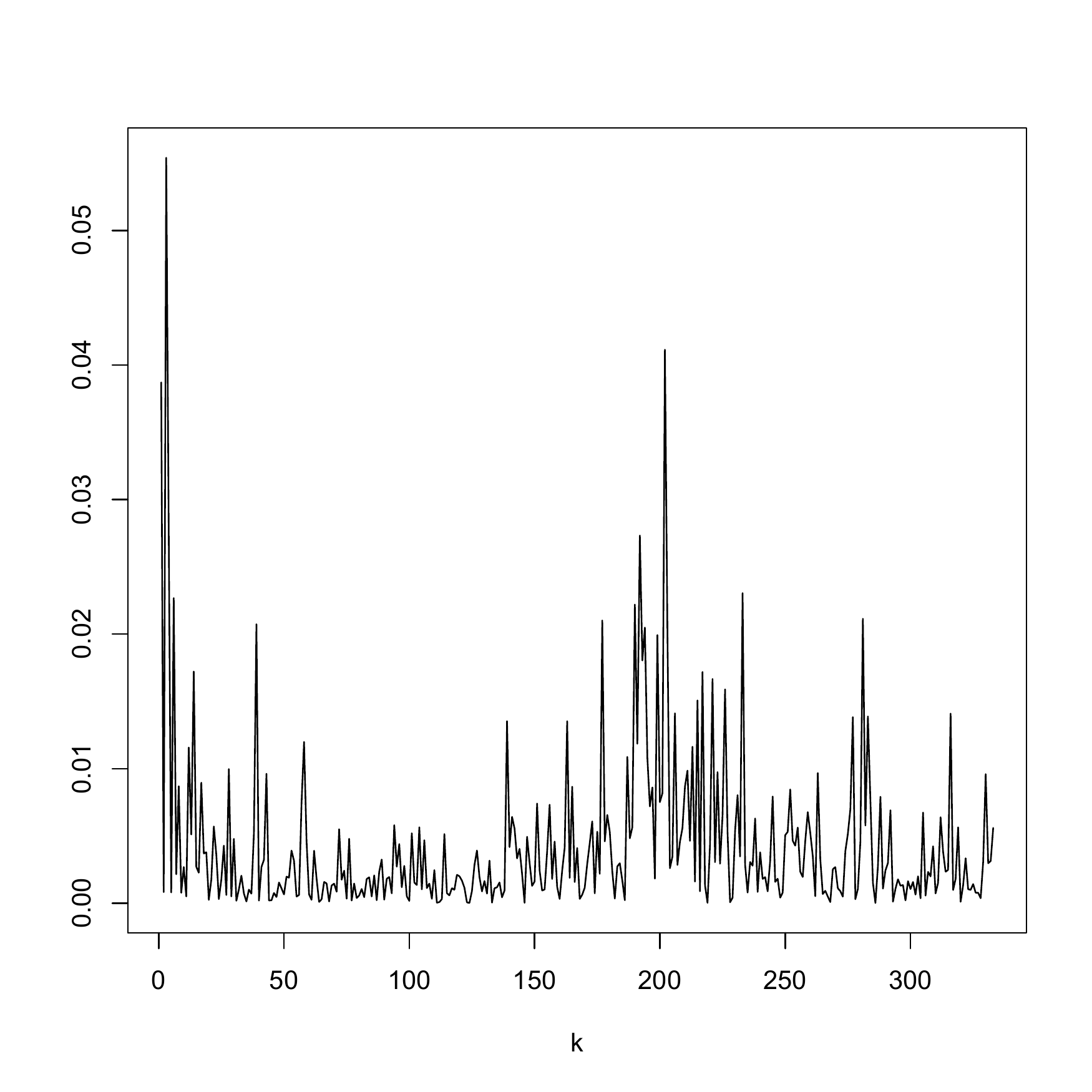}
  \label{fig:spdiff}
\end{subfigure}
\caption{Top: the sequence $\{|\tilde{X}_{s_k, e_k}^{b_k}|\}_{k=1}^{T-1}$ of Example 3.1; bottom: the sequence 
$\{Z_k\}_{k=1}^{333}$ of Example 3.1.
}
\label{fig:spdd}
\end{figure}

{\bf Example 3.1} \hspace{2pt}  As in Section \ref{sec:perfcomp} and Example 2.1, we consider the model $X_t = f_t + \varepsilon_t$,
$t = 1, \ldots, 1000$, where $f_t$ is the \verb+extreme.teeth+ signal.
However, the set-up in the current example is even more challenging in that we consider $\varepsilon_t$ i.i.d.
$\sim N(0, 0.4^2)$ as compared to $\varepsilon_t \sim N(0, 0.3^2)$ in Section \ref{sec:perfcomp} and Example 2.1.
For a particular realisation of this model, we compute the WBS2 solution path $\tilde{\mathcal P}$ with $\tilde{M} = 100$
and plot the resulting sequence $\{|\tilde{X}_{s_k, e_k}^{b_k}|\}_{k=1}^{T-1}$ (Figure \ref{fig:spdd}, top plot). Recall that by the construction of 
$\tilde{\mathcal P}$, this sequence is non-increasing.

The true number of change-points in $f_t$ is $N=199$. A visual inspection of the top plot in Figure
\ref{fig:spdd} reveals an increasing (negative) gradient in $\{|\tilde{X}_{s_k, e_k}^{b_k}|\}_{k=1}^{T-1}$
at around $k = 199$. This appears to align well with the assertion of part (iii) of Theorem \ref{th:solpath},
according to which, on a set of high probability, the elements of $\{|\tilde{X}_{s_k, e_k}^{b_k}|\}_{k=1}^{T-1}$ are of a higher order
of magnitude ($\gtrsim \underline{f}_T T^{1/2}$) for $k = 1, \ldots, N$ than they are for $k = N+1, \ldots, T-1$
($O(\log^{1/2}T)$). This observation provides a basis for our procedure for estimating $N$: our
$\hat{N}$ will be an estimator of the location of this change in orders of magnitude, or equivalently
the location of this large negative gradient.

In order to perform the estimation of this location, it is natural to work on the log scale: we denote
$Y_k = \log |\tilde{X}_{s_k, e_k}^{b_k}|$. The reason is that taking the differences $Z_k = Y_{k} - Y_{k+1}$
leads to the cancellation of the $\log\,T$ terms for $k = 1, \ldots, N-1$ and the cancellation 
of the $\log\,\log\,T$ terms for $k = N+1, \ldots, K$ (where $K$ is the largest $k$ for which  
$|\tilde{X}_{s_{k+1}, e_{k+1}}^{b_{k+1}}| \ge C \log^{1/2} T$ for a certain constant $C$). Therefore,
the terms $Z_k$ for $k = 1, \ldots, N-1, N+1, \ldots, K$ will be bounded in $T$, whereas the single term
$Z_N$ will tend to infinity with $T$ at the speed of $\log\,T$. As a consequence, it is tempting to consider setting $\hat{N}$
to the value of $k$ for which $Z_k$ is the largest. Our new model selection criterion does exactly
that, but with a certain finite-sample correction, which we motivate next.

The bottom plot of Figure \ref{fig:spdd} shows the sequence $Z_k$ for $k = 1, \ldots, K = 333$. In light
of the above discussion, it is unsurprising to see a prominent peak in $Z_k$ around $k = N = 199$ (in this particular
example, the location of this local peak is $k = 202$). Yet,
setting $\hat{N} = \arg\max_{k=1, \ldots, K} Z_k$ would have failed in this particular case, as the
global maximum of $Z_k$ is achieved for $k = 3$. However, one way of checking that $\hat{N} = 3$ leads to
an infeasible model in this example is to examine the size of $Y_4$, the smaller term entering the difference
$Z_3 = Y_3 - Y_4$. If $k = 3$ were to be the true model size, then by
part (iii) of Theorem \ref{th:solpath}
we would expect $|\tilde{X}_{s_4, e_4}^{b_4}| = \exp(Y_4)
= O(\log^{1/2}T)$. We use a threshold to decide whether $|\tilde{X}_{s_4, e_4}^{b_4}|$ is indeed of this magnitude, and
the form of this ``universal'' threshold is $1.2\, \hat{\sigma}_T \{2 \log\, T\}^{1/2}$ (where 
$\hat{\sigma}_T$ is the MAD estimate of $\sigma$ with $\{2^{-1/2}(X_{t+1} - X_t)\}_{t=1}^{T-1}$ on input);
this formula will be motivated in Sections \ref{sec:sdllth} and \ref{sec:prac}. In this particular example, we have
$|\tilde{X}_{s_4, e_4}^{b_4}| \approx 2.25 > 2.0 \approx 1.2\, \hat{\sigma}_T \{2 \log\, T\}^{1/2}$, and therefore
$k = 3$ is rejected as a feasible model. Our procedure then goes on to examine the next largest $Z_k$,
which in this case is $Z_{202}$; this leads to examination of $|\tilde{X}_{s_{203}, e_{203}}^{b_{203}}|$,
whose magnitude is well under 
$1.2\, \hat{\sigma}_T \{2 \log\, T\}^{1/2} \approx 2.0$. As a consequence, the model with 202 change-points is accepted
and we set $\hat{N} = 202$ (over-estimating the true number of change-points by 3). (End of example.)

\vspace{10pt}

We now comment heuristically on some aspects of Example 3.1 before introducing a formal algorithmic description of our
new ``Steepest Drop to Low Levels'' (SDLL) model selection procedure.

The name ``Steepest Drop to Low Levels'' has its origins in the fact that our model selection procedure does not
merely look for the ``steepest drop'' (= largest negative gradient) in $\log |\tilde{X}_{s_k, e_k}^{b_k}|$ (this occurs at
$k = 3$ in the above example) but it searches for the steepest drop followed by ``low levels'' (= values of $|\tilde{X}_{s_k, e_k}^{b_k}|$
that fall under the threshold); this occurs at $k = 202$ in the example.

We emphasise that the use of thresholding in the SDLL model selection is new and different from the classical use
of thresholding in change-point detection problems (as in e.g. \cite{f14a}). In classical thresholding,
a threshold is used as the main model selection device in the sense that its value is used to separate significant from 
insignificant change-point candidates. This happens in a  ``continuous'' manner, in the sense that even small changes in
the threshold value can lead to the exclusion or inclusion of additional candidate change-points and therefore
to changes in the selected model. A satisfactory threshold should in theory be linear with $\sigma$, so due to this
sensitivity of the selected model to the threshold value, the user should have the ability to estimate $\sigma$ accurately,
which is not always possible in frequent change-point settings, as illustrated earlier.

By contrast, SDLL uses thresholding only as a secondary model selection device, the main criterion being the size of the
negative gradient in $\log |\tilde{X}_{s_k, e_k}^{b_k}|$. In SDLL, thresholding is used ``discretely'' in the sense that the consecutive
models being tested via thresholding are often very remote from each other in terms of their numbers of change-points; in the
example above, we first tested a model with 3 change-points followed by a model with 202 change-points. In other words,
we used thresholding not so much to ``choose'' a model, but to test the feasibility of models pre-ordered according to
their respective negative gradients in $\log |\tilde{X}_{s_k, e_k}^{b_k}|$ (rather than according to their numbers of change-points).
The model with 3 change-points was rejected as easily as the model with 202 change-points was accepted.
This suggests that SDLL does not require the threshold to be selected as precisely as is the case in classical
thresholding. This is potentially a useful feature of SDLL as it 
facilitates accurate model selection in the presence
of frequent change-points, when the user is not always able to estimate $\sigma$ (and hence the threshold)
well.

We are now in a position to introduce a complete algorithm describing our SDLL model selection procedure.
It is defined by the WBS2.SDLL routine below, which takes three parameters on input: 
$\tilde{\mathcal P}$ is a WBS2 solution path, whose computation is outlined in Section \ref{sec:solpath};
$\tilde{\zeta}_T$ is a threshold; and $\beta \in (0, 1)$ is a constant.
In the remainder of this paper, we use $\tilde{\zeta}_T = \tilde{C} \hat{\sigma}_T \{ 2 \log\,T   \}^{1/2}$, and 
this choice will be motivated in Section \ref{sec:sdllth}; the choice of the constants $\tilde{C}$ and $\beta$ will be described in
Section \ref{sec:prac}.

\vspace{10pt}

\begin{algorithmic}
\Function{WBS2.SDLL}{$\tilde{\mathcal P}$, $\tilde{\zeta}_T$, $\beta$}
\If {$|\tilde{\mathcal P}| = 0$}
\State STOP
\EndIf
\If {$|\tilde{X}_{s_1, e_1}^{b_1}| < \tilde{\zeta}_T$}
\State $\hat{N} := 0$ 
\Else
\State $K := \max \{ k\,\,\,:\,\,\, |\tilde{X}_{s_{k+1}, e_{k+1}}^{b_{k+1}}| \ge \beta\,\tilde{\zeta}_T  \}$
\If {$K = 0$}
\State $\hat{N} := 1$
\Else
\State $Z_k := \log|\tilde{X}_{s_{k}, e_{k}}^{b_{k}}| - \log|\tilde{X}_{s_{k+1}, e_{k+1}}^{b_{k+1}}|$, $k = 1, \ldots, K$
\If {$\exists\,\,k = 1, \ldots, K\,\,\,\text{s.t.}\,\,\, |\tilde{X}_{s_{k+1}, e_{k+1}}^{b_{k+1}}| \le \tilde{\zeta}_T $}
\State $\hat{N} := \arg\{\max_{k=1,\ldots,K} Z_k\,\,\,\text{s.t.}\,\,\, |\tilde{X}_{s_{k+1}, e_{k+1}}^{b_{k+1}}| \le \tilde{\zeta}_T     \}$
\Else
\State $\hat{N} := K + 1$
\EndIf
\EndIf
\State $(\tilde{\eta}_1, \ldots, \tilde{\eta}_{\hat{N}}) := \textsc{Sort}((b_1, \ldots, b_{\hat{N}}), \textsc{Increasing})$
\EndIf
\EndFunction
\end{algorithmic}

If the magnitudes of $|\tilde{X}_{s_k, e_k}^{b_k}|$
all fall under $\tilde{\zeta}_T$, the procedure returns zero change-points. Otherwise, the procedure extracts all
those indices $k = 1, \ldots, K$ for which $|\tilde{X}_{s_{k+1}, e_{k+1}}^{b_{k+1}}| \ge \beta \tilde{\zeta}_T$; this is to ensure
that we only work with those $k$ for which $|\tilde{X}_{s_{k+1}, e_{k+1}}^{b_{k+1}}| \ge C\,\log^{1/2}T$ as motivated
in Example 3.1. Within this range of $k$, it then looks for the maximum negative gradient in $\log|\tilde{X}_{s_k, e_k}^{b_k}|$,
subject to the constraint that $|\tilde{X}_{s_{k+1}, e_{k+1}}^{b_{k+1}}| \le \tilde{\zeta}_T$, and assigns $\hat{N}$ to be that location;
if the constraint $|\tilde{X}_{s_{k+1}, e_{k+1}}^{b_{k+1}}| \le \tilde{\zeta}_T$ is never satisfied,
this necessarily means that the end of the range has been reached, in which case $\hat{N} = K+1$.

We now comment on a few important aspects of the SDLL model selection. As described earlier 
in the discussion of Example 3.1, SDLL is not a classical thresholding model selection procedure,
but also does not used a penalty-based approach. Its execution, not including the call to the $\textsc{Sort}$ routine,
is of computational order $O(T)$. Under the assumptions of Theorem \ref{th:sdll}, the execution of the 
$\textsc{Sort}$ routine is of computational order $O(1)$ with high probability. Unlike some penalty-based
approached, SDLL does not need to know the maximum number of change-points present in the data,
in either theory or practice.

\subsubsection{Theoretical properties of SDLL model selection}
\label{sec:sdllth}

In this section, we formulate a complete consistency theorem for the WBS2 method equipped with
the SDLL model selection criterion, as regards the estimation of both the number $N$ and the locations
$\eta_1, \ldots, \eta_N$ of the change-points. We have the following result.

\begin{theorem}
\label{th:sdll}
Let $X_t$ follow model (\ref{eq:model})
and let Assumption \ref{ass:tech} hold. Let $N$ and $\eta_1, \ldots, \eta_N$ denote,
respectively, the number and the locations of the change-points. Let $\tilde{\mathcal P}$ be the solution path of the WBS2 
algorithm, defined in Section \ref{sec:solpath}. Let $\hat{N}$ and $\tilde{\eta}_1, \ldots, \tilde{\eta}_{\hat{N}}$ be the estimates
of the number and the locations (respectively) of the change-points, returned by the SDLL model selection criterion as defined
by the WBS2.SDLL routine with $\tilde{\mathcal P}$, $\tilde{\zeta_T}$ and $\beta$ on input, with $\beta \in (0, 1)$ and
$\tilde{\zeta_T} = \tilde{C} \hat{\sigma}_T \{ 2 \log\,T   \}^{1/2}$ for a large enough constant $\tilde{C}$. 
Define the events
\begin{eqnarray*}
\Theta_{\theta, T} & = & \{(1+\theta)^{-1} < \hat{\sigma}_T/\sigma < 1+\theta\},\\
{\mathcal E}_T & = & \{  \hat{N} = N\quad \cap\quad \max_{i=1, \ldots, \hat{N}} |\eta_i - \tilde{\eta}_i| \le C_1(\Delta) (\underline{f}_T)^{-2} \log\,T \},
\end{eqnarray*}
where $\theta$ is a positive constant and $C_1(\Delta)$ is defined in the statement of Theorem \ref{th:solpath}. For $T$ large enough,
we have
\begin{equation}
\label{eq:lowbnd2}
P({\mathcal E}_T) \ge 1 - \alpha_T - \frac{1}{2}N(N+1) (2C_1(\Delta)(\underline{f}_T)^{-2}\log\,T + 1)^2 (1 - \delta^2/9)^{\tilde{M}} - P(\Theta_{\theta, T}^c),
\end{equation}
where $\alpha_T$ is defined in the statement of Theorem \ref{th:solpath}.
\end{theorem}

The remarks below complement and extend the discussion of Theorem \ref{th:solpath} in Section \ref{sec:spth}. In that discussion,
we already clarified that $\alpha_T \to 1$ as $T \to \infty$ and
that $\frac{1}{2}N(N+1) (2C_1(\Delta)(\underline{f}_T)^{-2}\log\,T + 1)^2 (1 - \delta^2/9)^{\tilde{M}} = O(T^{-1})$ if $\tilde{M} \ge C\,\log\,T$ for a large enough
constant $C$. We note that for $P(\Theta_{\theta, T}^c) \to 0$, we do not even require that $\hat{\sigma}_T$ should be a consistent
estimator of $\sigma$, but only that the ratio $\hat{\sigma}_T/\sigma$ be bounded with probability tending to 1 with $T$. This is emphasised
in order to reflect the empirical fact that $\sigma$ may be difficult to estimate well in frequent change-point scenarios.

Proving such a boundedness condition is typically not problematic for a variety of commonly used estimators of $\sigma$. As an example,
we now sketch how boundedness from above can be shown for a simplified Median Absolute Deviation estimator, defined by
\[
\hat{\sigma}_T^{\mathrm{MAD}} = C\, \mathrm{med}\{ |X_{t+1} - X_t|  \}_{t=1}^{T-1},
\]
where $C$ is a certain universal normalising constant and med is the median operator. Let avg denote the sample mean operator,
and let $a_t = C|X_{t+1} - X_t|$.
Similarly to \cite{m91}, recalling that the median measures the centre of the data in the $l_1$ sense, we have
\begin{eqnarray*}
|\mathrm{avg}\{a_t\} - \mathrm{med}\{a_t\}| & = & |\mathrm{avg}\{  a_t - \mathrm{med}\{a_t\}   \}| \le \mathrm{avg}|a_t - \mathrm{med}\{a_t\}|\\
& \le & \mathrm{avg}|a_t| = \mathrm{avg}\{a_t\},
\end{eqnarray*}
which gives $\mathrm{med}\{a_t\} \le 2\,\mathrm{avg}\{a_t\}$. Therefore the boundedness from above of $\hat{\sigma}_T^{\mathrm{MAD}}$
will follow from that of $\mathrm{avg}\{a_t\}$. For the latter, note that its expectation, $\mathbb{E}(\mathrm{avg}\{a_t\})$, is necessarily
bounded, regardless of the number of change-points in the signal, due 
to Assumption \ref{ass:tech}(c). Therefore, it suffices to establish a concentration bound on $\mathrm{avg}\{a_t\}$ around
its expectation, but this is a standard result, available, e.g., in Theorem 1.4 in \cite{b98}, which applies to zero-mean 
weakly-dependent processes satisfying Cramer's conditions, such as $a_t - \mathbb{E}(a_t)$. This completes the sketch of this argument.

Finally, note that the rate of $O(\log^{1/2}T)$ for the threshold $\tilde{\zeta_T}$ is standard and also appears, for example, in the standard
Wild Binary Segmentation algorithm of \cite{f14a} and in the Tail-Greedy bottom-up method of \cite{f18}.

\subsection{WBS2.SDLL vs other adaptive change-point model selection methods}
\label{sec:compad}

In this section, we compare and contrast WBS2.SDLL, in a more thorough way than was done in Section \ref{sec:intro},
to some existing methods in the literature which select the change-point model (i.e. estimate $N$) by
choosing the value of an appropriate tuning parameter in a data-driven way; for brevity, we refer to such techniques as adaptive.

In this sense, WBS2.SDLL can also be interpreted as an adaptive technique: given a sorted WBS2 solution path $\tilde{{\mathcal P}}$, constructed
as described in Section \ref{sec:solpath}, the SDLL model selection procedure regards as significant the $\hat{N}$ change-points that
correspond to the first $\hat{N}$ CUSUMs $|\tilde{X}_{s_k, e_k}^{b_k}|$, for $k = 1, \ldots, \hat{N}$, which are also the largest ones. Therefore
SDLL can be interpreted as an ``adaptive thresholding'' technique, which retains a change-point candidate $b_k$ if and only if
$|\tilde{X}_{s_k, e_k}^{b_k}| \ge \lambda_{SDLL}(X)$ and discards it otherwise, where $\lambda_{SDLL}(X)$ is a certain data-driven
threshold, chosen by the SDLL procedure from the shape of the solution path $\tilde{{\mathcal P}}$ as described in the definition of SDLL
in Section \ref{sec:sdllmet}.

\cite{bm01} and \cite{bm06} propose the adaptive choice of what they refer to as the minimal penalty, in the context of penalised Gaussian model selection in which the cost function
is the $L_2$ risk, via an algorithm referred to in later works as ``dimension jump''. The main idea of dimension jump is as follows. Consider
a family of penalties of the form $CD_m / T$, where $D_m$ is the dimensionality of the space of candidate models (in our context, this would be 
the postulated number of change-points). Heuristically, too small a choice of the constant $C$ when minimising the corresponding penalised $L_2$ risk criterion will lead 
to models with a lot of change-points, whereas a large (or indeed too large) a choice will reduce this considerably. Therefore the dimension jump 
approach looks at the graph of the function $C/\sigma^2 \mapsto D_{\hat{m}}$ and chooses the constant $C$ that corresponds to the largest jump in this
function, hoping that this will locate the ``sweet spot'' between penalties that are too small and those that are too large.

In the approach termed ``slope estimation" \citep{l05} the observation is that the optimal penalty for Gaussian model selection under the $L_2$ risk, in the sense
of \cite{bm06}, depends on the unknown parameter $\sigma^2$. For any postulated number of change-points $N'$, let $\hat{f}_t^{N'}$ be the best
fit to the data in the $L_2$ sense that contains exactly $N'$ change-points. For $N'$ large enough, the function
$N' \mapsto \sum_{t=1}^T (X_t - \hat{f}_t^{N'})^2$ is approximately linear with slope $-\sigma^2 / T$, which can in principle be used to estimate
$\sigma^2$, which can then be plugged into the optimal penalty.

Specific algorithmic ideas regarding the implementation of dimension jump and slope estimation in the context of the multiple
change-point problem studied in this paper appear, amongst others, in \cite{l05} and \cite{bmm12}. Our comparative simulation study in Section 
\ref{sec:simstu} includes both these approaches and describes the details of the R packages used.
 \cite{a19} is an excellent and comprehensive review
article on minimal penalties and the slope heuristics, which describes these two approaches and their variants in detail,
including an unpublished two-stage refinement by Rozenholc, branded ``statistical base jumping".

We note that SDLL is fundamentally different from both dimension jump and slope estimation. To start with, SDLL makes its model 
selection decision based on the shape of the function $k \mapsto \log\{ |\tilde{X}_{s_k, e_k}^{b_k}|\}$, where $|\tilde{X}_{s_k, e_k}^{b_k}|$
are the WBS2 CUSUMs sorted in decreasing order. As described above, dimension jumps considers the shape of an entirely different function,
which does not appear obviously related to that considered by SDLL. Slope estimation considers the tail behaviour (which is very different
from the steepest drop to low levels as considered by SDLL) of the function $N' \mapsto \sum_{t=1}^T (X_t - \hat{f}_t^{N'})^2$, which for
each $N'$ measures the impact of the $N'$ most important change-points in the data. Therefore if we were to take the difference 
$\sum_{t=1}^T (X_t - \hat{f}_t^{N'})^2 - \sum_{t=1}^T (X_t - \hat{f}_t^{N'+1})^2$, this would provide a measure of the importance of the 
$N'+1$st change-point. Even this is different from the measure of importance of each change-point used by SDLL, which is the WBS2 CUSUM. The attractive feature
of considering WBS2 CUSUMs in the context of SDLL, rather than, say, the differences
$\sum_{t=1}^T (X_t - \hat{f}_t^{N'})^2 - \sum_{t=1}^T (X_t - \hat{f}_t^{N'+1})^2$, is that the WBS2 procedure targets and returns the {\em maximum}
available CUSUM at each stage. As a result, it is hoped that WBS2 maximises the measure of each significant change-point, without unduly
inflating the CUSUMs corresponding to the insignificant change-points, as these are uniformly bounded. This aggressive separation makes
it easier for SDLL to identify the drop in the sorted sequence of WBS2 CUSUMs. Indeed, in experiments not reported in this paper, we tried using
the SDLL criterion applied to measures of change-point significance closely related to $\sum_{t=1}^T (X_t - \hat{f}_t^{N'})^2 - \sum_{t=1}^T (X_t - \hat{f}_t^{N'+1})^2$ 
and the results were much less encouraging than those for WBS2.SDLL. For the same reason, SDLL does not appear to work well when coupled with the
standard binary segmentation.

While dimension jump and slope estimation are approaches to choosing a suitable penalty constant (which is then
used for model choice), SDLL is a direct model selector. In this latter category is also the proposal
of \cite{l05a}, who (motivated by the problem of selecting a suitable penalty constant from the data),
proposes a heuristic algorithm for estimating the number of change-points. The proposed estimator of
$N$ is the largest value of $N'$ for which the second derivative of the (normalised) empirical risk
$\sum_{t=1}^T (X_t - \hat{f}_t^{N'})^2$ is greater than a certain threshold. Other that the difference in how this 
approach and WBS2.SDLL measure the importance of each change-point (see the discussion in the previous
paragraph), the two model selectors: SDLL and \cite{l05a} use fundamentally different functionals of their
corresponding solution paths, with \cite{l05a} (in contrast to SDLL) requiring the provision of the maximum number of
change-points and a threshold parameter whose value appears critical to the success of the
procedure; the theoretical properties of the method in \cite{l05a} are not investigated.

\section{Practicalities and comparative simulation study}
\label{sec:num}

\subsection{Practicalities}
\label{sec:prac}

This section discusses our default choices and recommendations for the parameters of the WBS2 solution 
path procedure with the SDLL model selection; as before, we refer to this combination as WBS2.SDLL.

\vspace{10pt}

{\em The number $\tilde{M}$ of interval draws.} $\tilde{M}$, the maximum number of random interval draws in
each recursive execution of the \textsc{WBS2.Sol.Path} routine defined in Section \ref{sec:solpath}, is the only parameter of the WBS2 solution path.
In all numerical examples in this paper, we use $\tilde{M} = 100$. This value has not been optimised and it is likely
that other values may lead to even better performance.

{\em The estimator $\hat{\sigma}_T$ of $\sigma$.} We use the MAD estimator for Gaussian data, with $\{2^{-1/2}(X_{t+1} - X_t)\}_{t=1}^{T+1}$
on input. In our experience, this estimator generally behaves well in infrequent and moderate number of change-point scenarios.
Section \ref{sec:modsel} shows that it may not be a very accurate estimator of $\sigma$ in frequent change-point settings,
but this is of little relevance as Theorem \ref{th:sdll} shows that we only require an estimator for which
$\hat{\sigma}_T/\sigma$ is bounded (which even allows for inconsistent estimators).

{\em The threshold constant $\tilde{C}$.} Recall that the threshold $\tilde{\zeta_T}$, used in the SDLL model selection procedure,
is of the form $\tilde{C} \hat{\sigma}_T \{ 2 \log\,T   \}^{1/2}$. We numerically calibrate the multiplicative threshold constant $\tilde{C}$
over a range of sample sizes $T$, so that the WBS2.SDLL procedure correctly estimates zero change-points for signals that have none,
in either 90\% or 95\% of the simulated sample paths, based on 1000 simulations. For the remaining sample sizes within the range of $T$ considered,
we extrapolate $\tilde{C}$ linearly; for any sample sizes outside the range of $T$ considered, we extrapolate $\tilde{C}$ as continuous
and constant. For the 90\% level, the values of $\tilde{C}$ range from 1.42 for $T=10$ and under, to 1.135 for $T = 10000$ and above. For the 90\%
level, they range from 1.55 for $T = 10$ and under to 1.17 for $T = 10000$ and above. We refer to the two versions of the procedure as,
respectively, WBS2.SDLL(0.9) and WBS2.SDLL(0.95). To obtain $\tilde{C}$ for any $T$, we refer the reader to
our R code at \url{https://github.com/pfryz/wild-binary-segmentation-2.0}.

{\em Constant $\beta$.} Recall from the definition of the WBS2.SDLL routine in Section \ref{sec:sdll} that SDLL performs model selection
over those model sizes for which the corresponding WBS2 absolute CUSUM statistics are of magnitude $\beta \tilde{\zeta}_T$ or larger. We have tested
$\beta$ is the region $0.3$--$0.6$ and have found that the SDLL model choice is relatively robust to the values of $\beta$ within this range.
We use $\beta = 0.3$ in all examples of this paper and this is also the default value in our code.

{\em Reducing the randomness of the output of WBS2.SDLL.} WBS2 is a random procedure, in the sense that it relies on the randomly drawn intervals
$[s_m, e_m]$. As a result, the output of the WBS2.SDLL procedure can vary across executions on the same dataset. This variation can be particularly
pronounced for datasets for which the signal to noise ratio is very low.
Two naive ways to reduce or eliminate the randomness of the output of WBS2.SDLL might be: to set a fixed random seed before the execution of the procedure, or
to use a fixed grid for the interval draws. Fixing the seed is risky as the user has no knowledge of what particular seeds may lead to favourable
interval draws. On the other hand, using a fixed uniform grid makes it impossible to request any given number $\tilde{M}$ of intervals as this is constrained by the spacing of the 
grid. Also, this normally excludes short intervals (which can be good for detecting finer-scale change-points) unless $\tilde{M}$ is very large, which tends not to be the case in WBS2.

Instead, we propose the following solution:
\begin{enumerate}
\item
run the WBS2.SDLL procedure $R$ times;
\item
choose the run that leads to the median number of change-points across the $R$ runs;
\item
return this ``median run" along with the ensemble of estimated change-point locations pooled across all $R$ runs.
\end{enumerate}

Our experiments (with $R = 9$), not reported in this paper, suggest that the median run is remarkably stable with respect to the number and the locations of estimated change-points, and tends to show only very little, if any, randomness. Also, the pooled ensemble of change-point estimates can be used to construct the histogram of all change-point locations detected across the $R$ runs, which provides an interesting informal visualisation of the significance of the estimated change-point locations. We provide an implementation of this procedure in our R code
at \url{https://github.com/pfryz/wild-binary-segmentation-2.0}. However, to save computation time, in the remainder of the paper we consider the standard WBS2.SDLL procedure, without this extra randomness-reduction feature.

\subsection{Comparative simulation study}
\label{sec:simstu}

In the first part of the simulation study, we exhibit the finite-sample performance of the WBS2.SDLL 
method on signals with small/moderate numbers of change-points, and compare it with that of the competitors defined
in Section \ref{sec:perfcomp}.
Our test models are defined in 
Appendix B of \cite{f14a} and are as follows:
(1) the \verb+blocks+ model (length 2048, 11 change-points),
(2) the \verb+fms+ model (length 497, 6 change-points),
(3) the \verb+mix+ model (length 560, 13 change-points),
(4) the \verb+teeth10+ model (length 140, 13 change-points),
(5) the \verb+stairs10+ model (length 150, 14 change-points).
For techniques that require the specification of the maximum number of change-points, we set this to the
(rounded) length of the input signal divided by three; in all our examples, the true number of change-points is (well) within
the interior of this range. 
We set the random seed to 1 before running
100 simulations for each of the test signals and each of the competitors.

The average absolute error in estimating the number $N$ of change-points, denoted
by $\hat{E}|\hat{N} - N|$, is an easy-to-interpret error measure which, in our experience, tends to
correlate strongly with many measures of accuracy of the estimated change-point locations. In particular,
for each method, let $\hat{f}_t$ be the piecewise-constant function whose value between
each pair of consecutive estimated change-points is the mean of the data over the 
interval delimited by this pair of change-points, and let $\hat{E}(\hat{f} - f)^2$ be the mean-square error in
estimating $f_t$, averaged across the simulations. We have found that the empirical across-method correlation
between $\log\{\hat{E}|\hat{N} - N|\}$ and $\log\{\hat{E}(\hat{f} - f)^2\}$, averaged over the five
models tested, is 0.799.

\begin{table}[h]
\centering
\begin{tabular}{|c|c|c|c|c|c|}
  \hline
method & (1) & (2) & (3) & (4) & (5) \\
\hline
PELT+mBIC & 0  &  1  &  0  &   0  &   1\\
PELT+BIC & 1  &   1 &   0 &   0  &  1\\
MOSUM &  1  &  1  &  0 &   0  &  1\\
ID &  1  &   1  &  0  &   1  &  1\\
FDRSeg & 1  &  1 &   0  &  0  &  1\\
S3IB & 0 &   0  &  0  &  0  &  0\\
SMUCE & 0  &  1  &  0 &   0  &  0\\
CUMSEG & 0  &   0  &  0  &  0  &  1\\
FPOP & 1  &  1  &  0 &   0  &  1\\
DDSE & 1  &  1  &  1  &  1  &  1\\
DJUMP & 1  &  1 &   1 &   0  &  1\\
TGUH &  1 &   1 &   0  &  1  &  1\\
WBS-C1.0 &  1 &   0  &  1  &  1  &  1\\
WBS-C1.3 &  0  &  1  &  0 &   0  &  1\\
WBS-BIC & 1 &   1  &   0  &   1  &   1\\
WBS2.SDLL(0.9) & 1  &  1  &  0  &  1  &  1\\
WBS2.SDLL(0.95) &   1  &  1  &  0  &  1  &  1\\
\hline
\end{tabular}
\caption{Logical flags indicating the performance of each method on each of the models (1)--(5): 1 means
$\hat{E}|\hat{N} - N| < 1$ (the true number of change-points is mis-estimated by less than 1 on average
across 100 simulated sample paths); 0 means $\hat{E}|\hat{N} - N| \ge 1$.\label{tab1}}
\end{table}

Table \ref{tab1} shows the (method, model) combinations for which $\hat{E}|\hat{N} - N| < 1$, i.e.
the true number of change-points is mis-estimated by less than 1 on average. The WBS2.SDLL
method
is one of the better performers, achieving 
$\hat{E}|\hat{N} - N| < 1$ for four out of the five test models. The only exception is the \verb+mix+
model, the most challenging out of the five models tested, 
for which WBS2.SDLL(0.9) and WBS2.SDLL(0.95) achieve $\hat{E}|\hat{N} - N| = 1.41$ and 
$1.4$, respectively.

The other techniques achieving $\hat{E}|\hat{N} - N| < 1$ for four out of the five test models are
ID, DJUMP, TGUH, WBS-C1.0 and WBS-BIC, with DDSE being the only method achieving
$\hat{E}|\hat{N} - N| < 1$ for all five test models.

However, in the second part of the simulation study, we show that the impressive performance of 
DDSE does not appear to extend to signals with frequent change-points. We first provide,
in Table \ref{tab2}, the full
simulation results for the \verb+extreme.teeth+ example from Section \ref{sec:perfcomp}, for which
$N = 199$. The two WBS2.SDLL methods are clear winners; relative to them, the large value of 
$\hat{E}(\hat{N} - N)^2$ for DDSE originates from the fact that DDSE estimates $\hat{N} = 0$ for two
out of 100 sample paths. All other methods perform consistently poorly.

\begin{table}[h]
\centering
\begin{tabular}{|c|c|c|c|c|c|}
  \hline
method & $\hat{E}(\hat{N}) - N$ & $\hat{E}|\hat{N} - N|$ & $\hat{E}(\hat{N} - N)^2$ & $\hat{E}(\hat{f} - f)^2$ & time\\
\hline
PELT+mBIC & -198.97 & 198.97 & 39589.09 & 0.250 & 0.001\\
PELT+BIC & -193.23 & 193.23 & 37392.13 & 0.247 & 0.001\\
MOSUM &  -199.00 & 199.00 & 39601.00 & 0.250 & 0.007\\
ID &  -186.00 & 186.00 & 35842.88 & 0.239 & 0.039\\
FDRSeg & -139.32 & 139.32 & 22568.66 & 0.197 & 1.329\\
S3IB & -199.00 & 199.00 & 39601.00 & 0.250 & 1.738\\
SMUCE & -192.47 & 192.47 & 37057.27 & 0.251 & 0.006\\
CUMSEG & -199.00 & 199.00 & 39601.00 & 0.250 & 1.282\\
FPOP & -193.23 & 193.23 & 37392.13 & 0.247 & 0.001\\
DDSE & -3.71 &  6.49 &   803.19 & 0.051 & 0.721\\
DJUMP & -199.00 & 199.00 & 39601.00 & 0.250 & 0.069\\
TGUH &  -132.38 & 132.38 & 18543.98 & 0.194 & 0.061\\
WBS-C1.0 &  -160.84 & 160.84 & 25935.10 & 0.242 & 0.083\\
WBS-C1.3 &  -193.70 & 193.70 & 37531.66 & 0.250 & 0.083\\
WBS-BIC & -199.00 & 199.00 & 39601.00 & 0.250 & 0.517\\
WBS2.SDLL(0.9) & -0.52 &   3.52  &  26.42 & 0.049 & 0.193\\
WBS2.SDLL(0.95) &   -0.08 &   3.22  &  17.20 & 0.049 & 0.191\\
\hline
\end{tabular}
\caption{Various measures of performance for the competing methods on the 
{\tt extreme.teeth} example with $\sigma = 0.3$, averaged over 100 simulated
sample paths. ``Time'' denotes the average execution time in seconds (in R, on a 2015
iMac); the other column headings are self-explanatory.
\label{tab2}}
\end{table}

\begin{figure}[h]
\begin{subfigure}{0.5\textwidth}
  \centering
  \includegraphics[width=.9\linewidth]{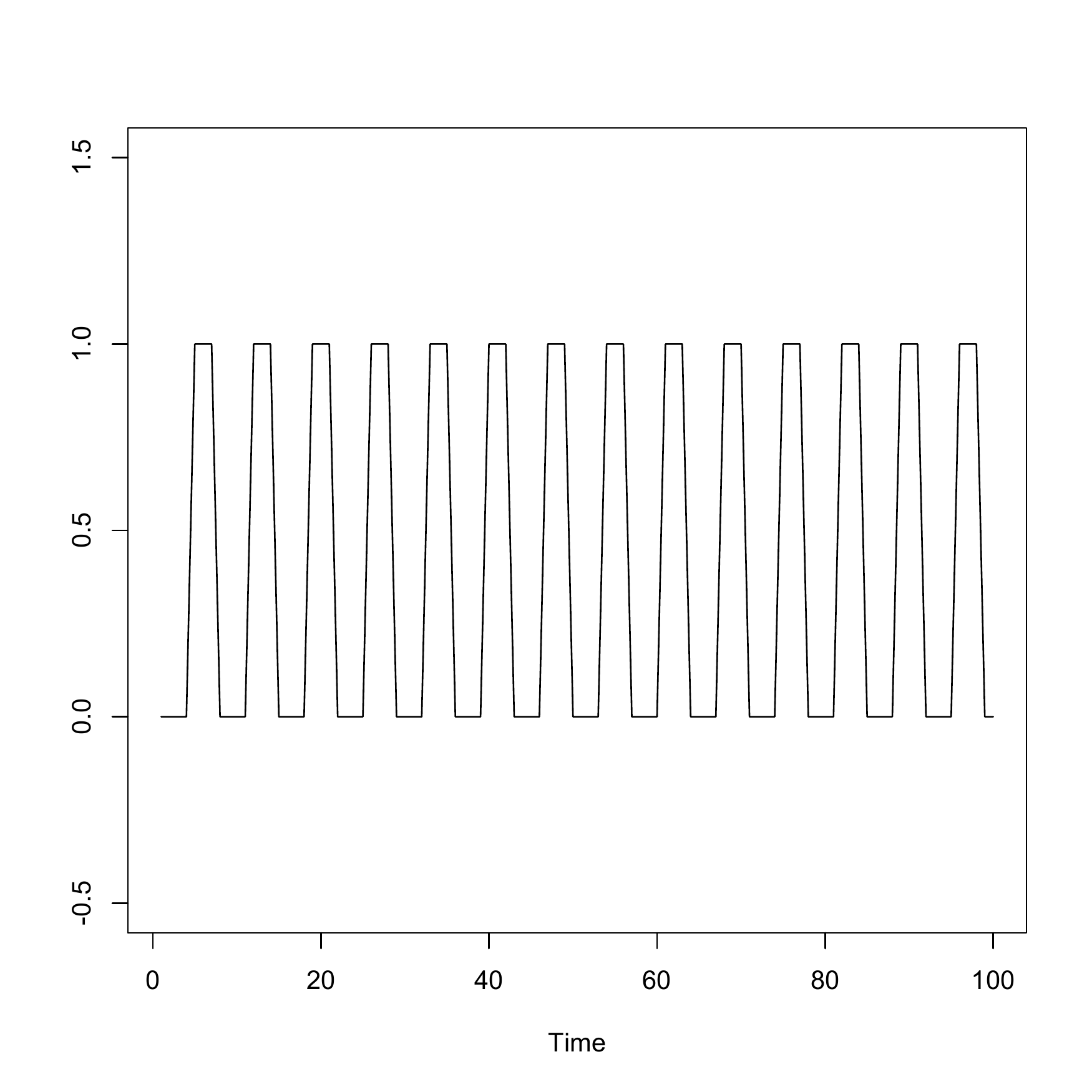}
\end{subfigure}
\begin{subfigure}{0.5\textwidth}
  \centering
  \includegraphics[width=.9\linewidth]{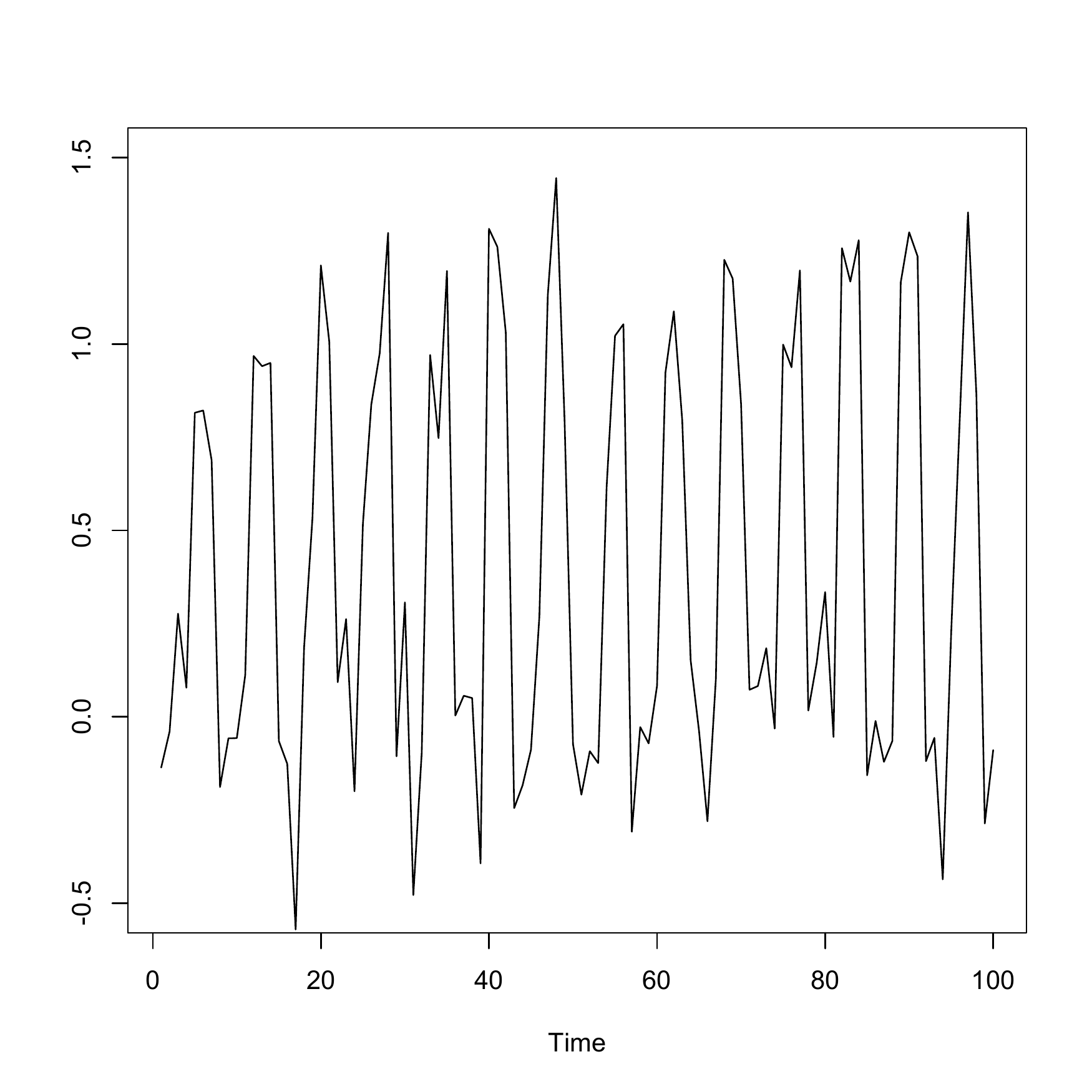}
\end{subfigure}
\caption{The first 100 observations of the {\tt extreme.extreme.teeth} signal (which continues in the same manner for $t = 1, \ldots, 700$);
left: with no noise; right: with additive i.i.d. Gaussian noise with mean zero and $\sigma = 0.2$.}
\label{fig:ext_ext_t}
\end{figure}

We further consider a slightly altered set-up in which the signal under consideration
has more frequent change-points, but less noise. The signal, labelled \verb+extreme.extreme.teeth+,
is one in which the pattern $0, 0, 0, 0, 1, 1, 1$ repeats itself 100 times, so that we have $T = 700$ and
$N = 199$. We use $\varepsilon_t \sim N(0, 0.2^2)$ (rather than the 
$\varepsilon_t \sim N(0, 0.3^2)$ used in the \verb+extreme.teeth+ example). See Figure \ref{fig:ext_ext_t}
for an illustration. The full simulation results are in Table \ref{tab3}. This time the two 
WBS2.SDLL methods are the only ones with acceptable performance and all the other methods, along
with DDSE, fail somewhat dramatically.

We close with a few further comparative remarks regarding DDSE and WBS2.SDLL.
Unlike WBS2.SDLL, the DDSE method requires the specification of the maximum number of change-points
and this was set to $[T/3]$, but we also tried $T-1$ and the results did not improve. Importantly, DDSE
appears to scale poorly: the execution on a signal of length $10^4$ took 22 seconds, was interrupted by
us on a signal of length $5 \times 10^4$ after several minutes, and we were unable to execute DDSE on 
a signal of length $10^5$ because of what we believe were memory-related issues. By contrast, the 
corresponding execution times for WBS2.SDLL were: 3.7 sec ($T = 10^4$),
15 sec ($T = 5 \times 10^4$), 30 sec ($T = 10^5$), which suggests empirical computation times close
to $O(T)$.

\begin{table}[h]
\centering
\begin{tabular}{|c|c|c|c|c|c|}
  \hline
method & $\hat{E}(\hat{N}) - N$ & $\hat{E}|\hat{N} - N|$ & $\hat{E}(\hat{N} - N)^2$ & $\hat{E}(\hat{f} - f)^2$ & time\\
\hline
PELT+mBIC & -198.97 & 198.97 & 39589.09 & 0.245 & 0.001\\
PELT+BIC & -189.95 & 189.95 & 36238.03 & 0.237 & 0.001\\
MOSUM &  -199.00 & 199.00 & 39601.00 & 0.245 & 0.007\\
ID &  -173.64 & 173.64 & 32457.76 & 0.220 & 0.026\\
FDRSeg & -140.77 & 140.77 & 24744.07 & 0.181 & 0.460\\
S3IB & -199.00 & 199.00 & 39601.00 & 0.245 & 0.865\\
SMUCE & -195.42 & 195.42 & 38197.30 & 0.245 & 0.006\\
CUMSEG & -199.00 & 199.00 & 39601.00 & 0.245 & 0.554\\
FPOP & -189.95 & 189.95 & 36238.03 & 0.237 & 0.001\\
DDSE & -167.16 & 167.28 & 33265.02 & 0.208 & 0.439\\
DJUMP & -199.00 & 199.00 & 39601.00 & 0.245 & 0.033\\
TGUH &  -84.92 &  84.92 &  8692.78 & 0.124 & 0.048\\
WBS-C1.0 &  -154.47 & 154.47 & 23939.55 & 0.230 & 0.068\\
WBS-C1.3 &  -192.37 & 192.37 & 37027.93 & 0.244 & 0.068\\
WBS-BIC & -199.00 & 199.00 & 39601.00 & 0.245 & 0.438\\
WBS2.SDLL(0.9) & 0.26 &  0.76  &    1.92 & 0.017 & 0.117\\
WBS2.SDLL(0.95) &   0.31 &  0.71  &   1.71 & 0.017 & 0.116\\
\hline
\end{tabular}
\caption{Various measures of performance for the competing methods on the 
{\tt extreme.extreme.teeth} example with $\sigma = 0.2$, averaged over 100 simulated
sample paths. ``Time'' denotes the average execution time in seconds (in R, on a 2015
iMac); the other column headings are self-explanatory.
\label{tab3}}
\end{table}

\subsection{Performance for heavy-tailed noise}

In this section, we illustrate the performance of WBS2.SDLL in the presence of heavy-tailed
noise. The noise distribution is not known to the algorithm, and it proceeds as it would if the noise were
Gaussian. We consider the \verb+extreme.teeth+ signal and use the following noise distributions:
$t_{2.5}$, $t_5$, $t_7$, $t_{10}$ and Gaussian, all scaled, as in Section \ref{sec:perfcomp}, to have the standard deviation
of $0.3$.

\begin{figure}[h]
\begin{subfigure}{\textwidth}
  \centering
  \includegraphics[width=.8\linewidth, height = .28\textheight]{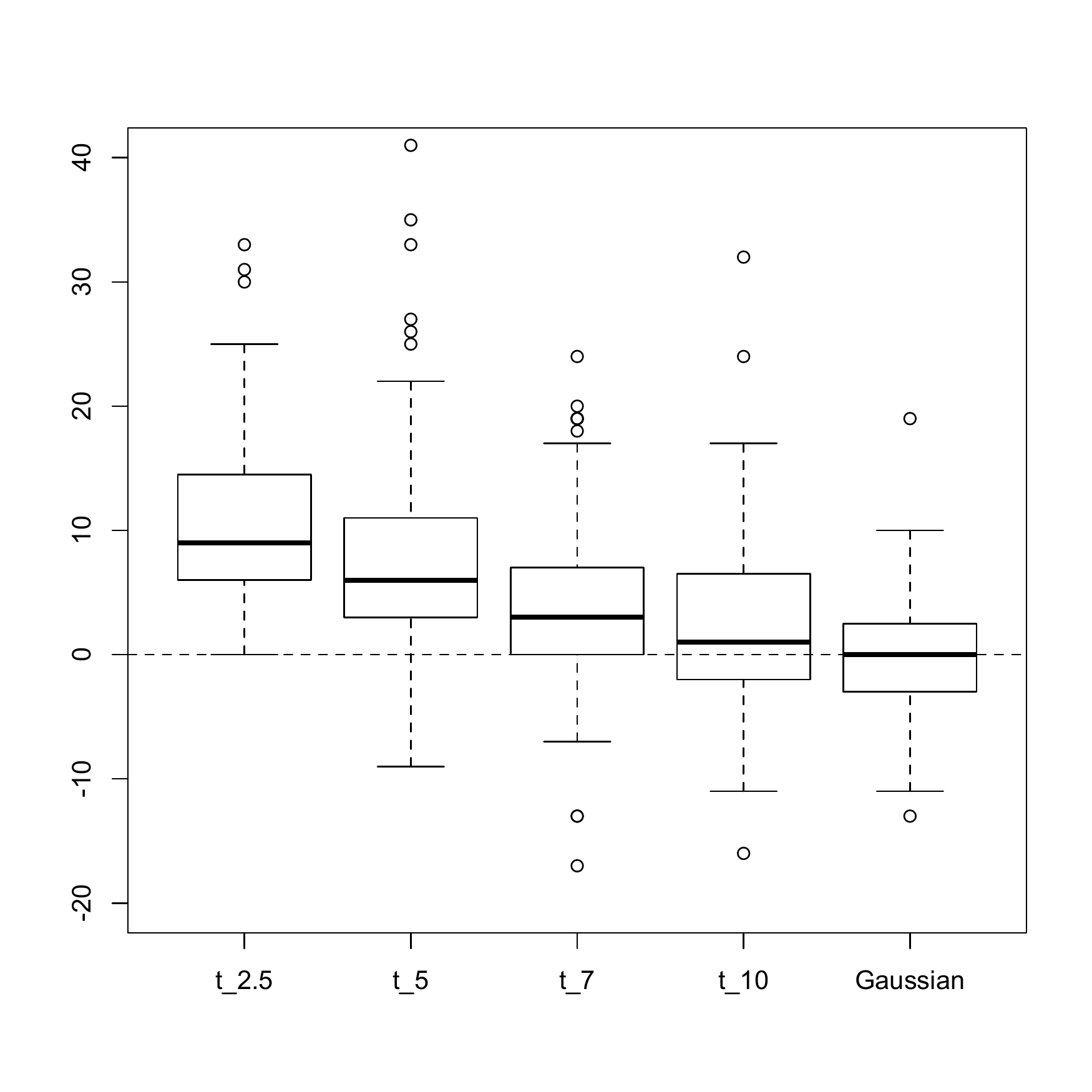}
  \label{fig:ht90}
\end{subfigure}\\
\begin{subfigure}{\textwidth}
  \centering
  \includegraphics[width=.8\linewidth, height = .28\textheight]{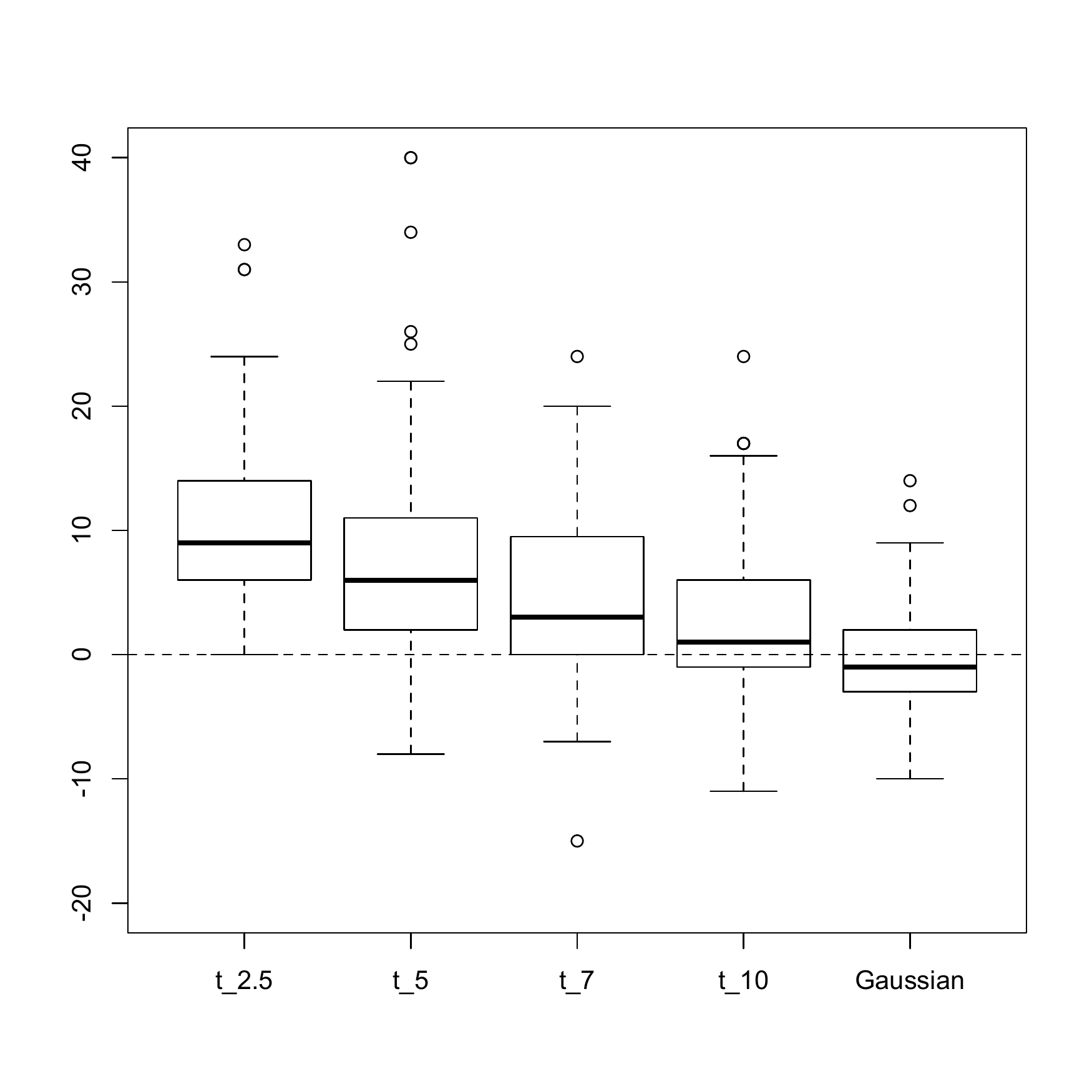}
  \label{fig:ht95}
\end{subfigure}
\caption{From top to bottom and left to right: boxplots of $\hat{N} - N$ over 100 simulated sample paths, for the noisy {\tt extreme.teeth} signal,
for WBS2.SDLL(0.9) [top] and WBS2.SDLL(0.95) [bottom] and the following noise distributions, each with standard deviation 0.3: 
$t_{2.5}$, $t_5$, $t_7$, $t_{10}$ and Gaussian.
}
\label{fig:ht}
\end{figure}

Figure \ref{fig:ht} shows the results. We are encouraged by the fact that the upward bias in the estimation of $N$ appears to be rather small
even in the case of $t_{2.5}$, which is strongly heavy-tailed.

\section{London House Price Index}
\label{sec:ukhpi}

We consider monthly percentage changes in the UK House Price Index (UKHPI), for all property types,
in the 32 London boroughs (the 32 local authority districts that make up the Greater London county).
We exclude the City of London as its resident population is much smaller than in any of the 32 boroughs
and, as a result, the monthly fluctuations in the UKHPI for the City of London are much more variable.
As of December 2018, the data are available from \url{http://landregistry.data.gov.uk/app/ukhpi}. The 32-dimensional
time series $X_t = (X_t^{(1)}, \ldots, X_t^{(32)})'$ runs from February 1995 to September 2018, so that
$t = 1, \ldots, T = 284$. In the order of the indexing ($i = 1, \ldots, 32$), the boroughs are:
Barking and Dagenham,
Barnet,
Bexley,
Brent,
Bromley,
Camden,
City of Westminster,
Croydon,
Ealing,
Enfield,
Greenwich,
Hackney,
Hammersmith and Fulham,
Haringey,
Harrow,
Havering,
Hillingdon,
Hounslow,
Islington,
Kensington and Chelsea,
Kingston upon Thames,
Lambeth,
Lewisham,
Merton,
Newham,
Redbridge,
Richmond upon Thames,
Southwark,
Sutton,
Tower Hamlets,
Waltham Forest, and
Wandsworth.

\begin{figure}[h]
\begin{subfigure}{\textwidth}
  \centering
  \includegraphics[width=.8\linewidth, height = .28\textheight]{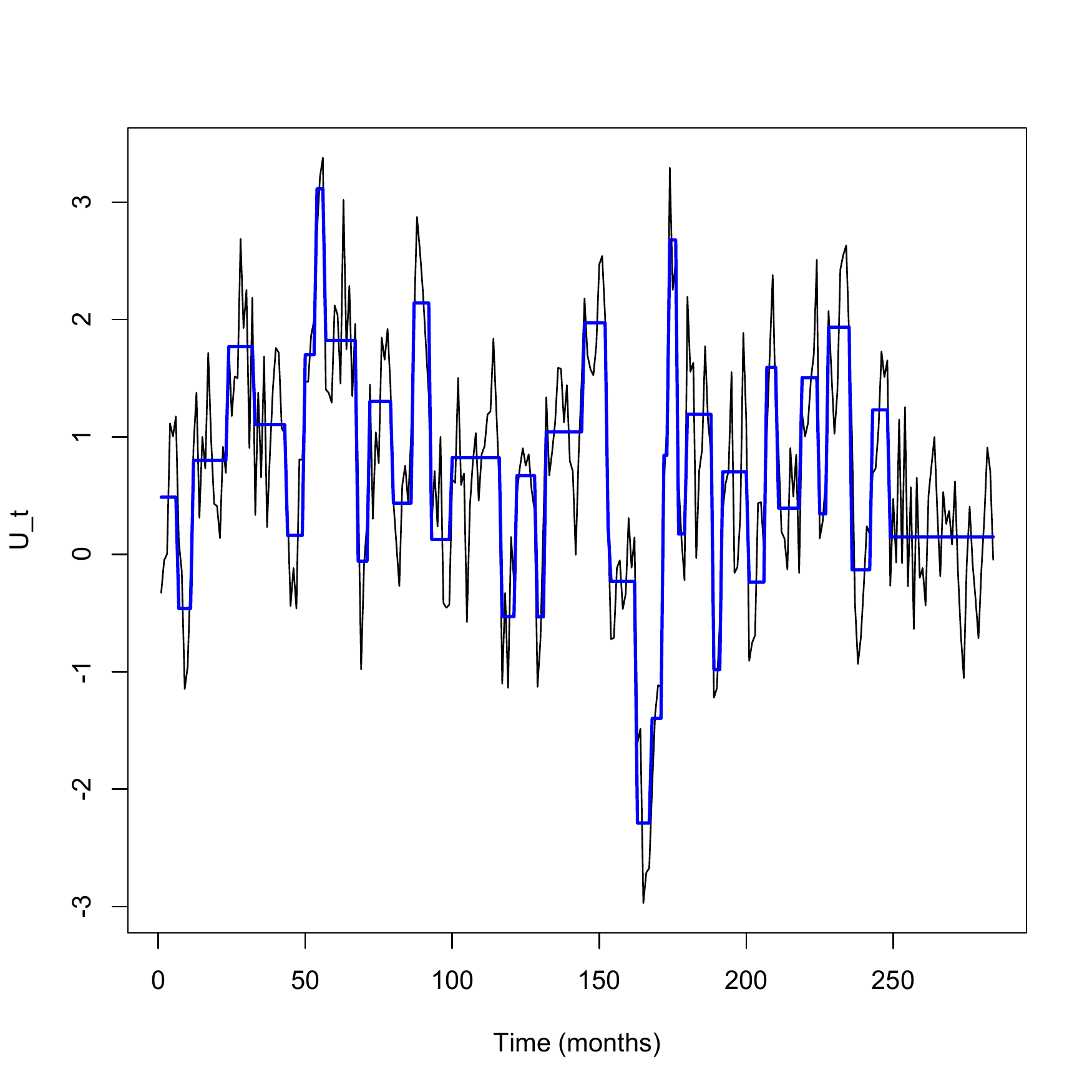}
  \label{fig:ut}
\end{subfigure}\\
\begin{subfigure}{\textwidth}
  \centering
  \includegraphics[width=.8\linewidth, height = .28\textheight]{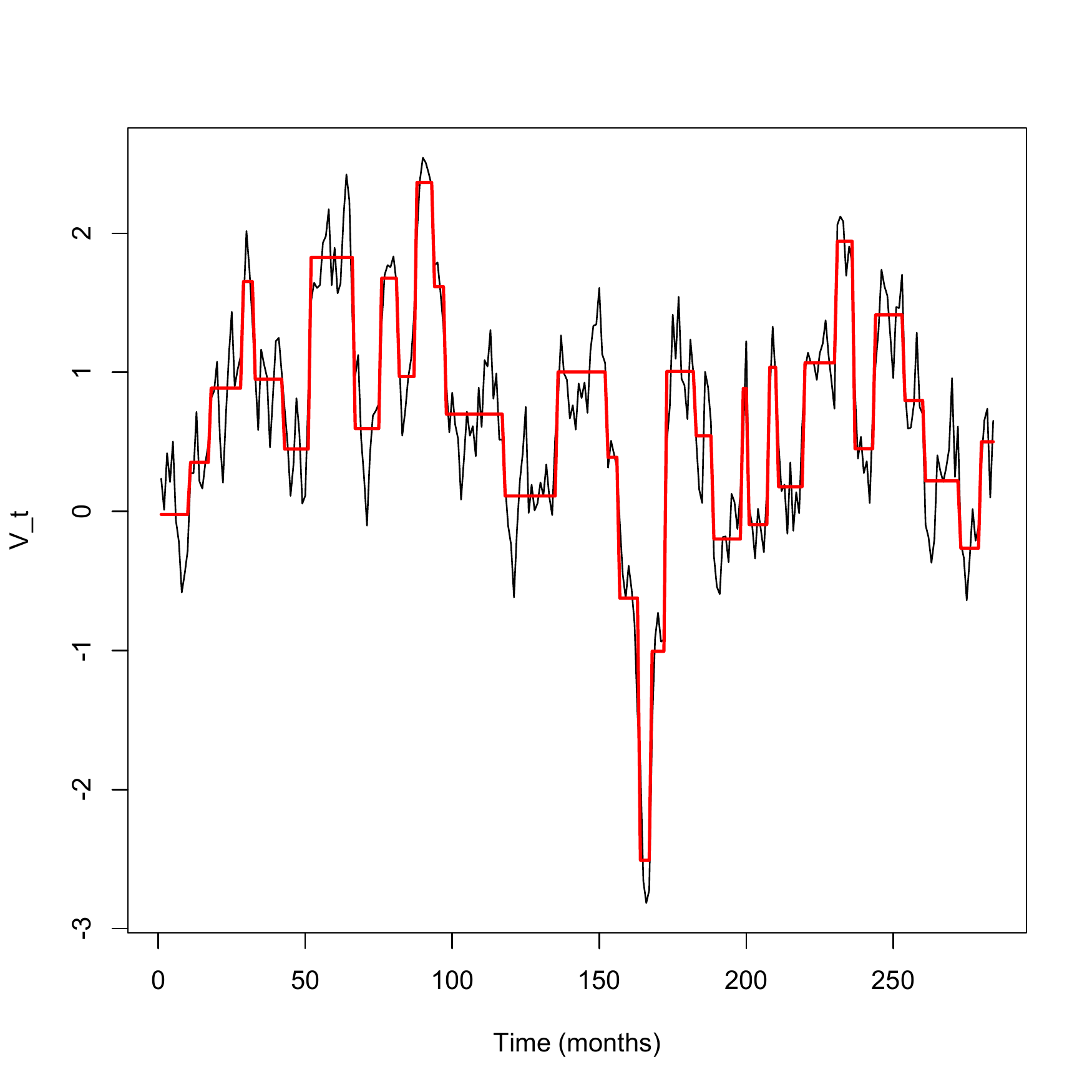}
  \label{fig:vt}
\end{subfigure}
\caption{Top: $U_t$ and $\hat{E}(U_t)$ (blue); bottom: $V_t$ and $\hat{E}(V_t)$ (red).
}
\label{fig:uv}
\end{figure}

\begin{figure}[p]
\begin{subfigure}{0.5\textwidth}
  \centering
  \includegraphics[width=.9\linewidth, height=.25\textheight]{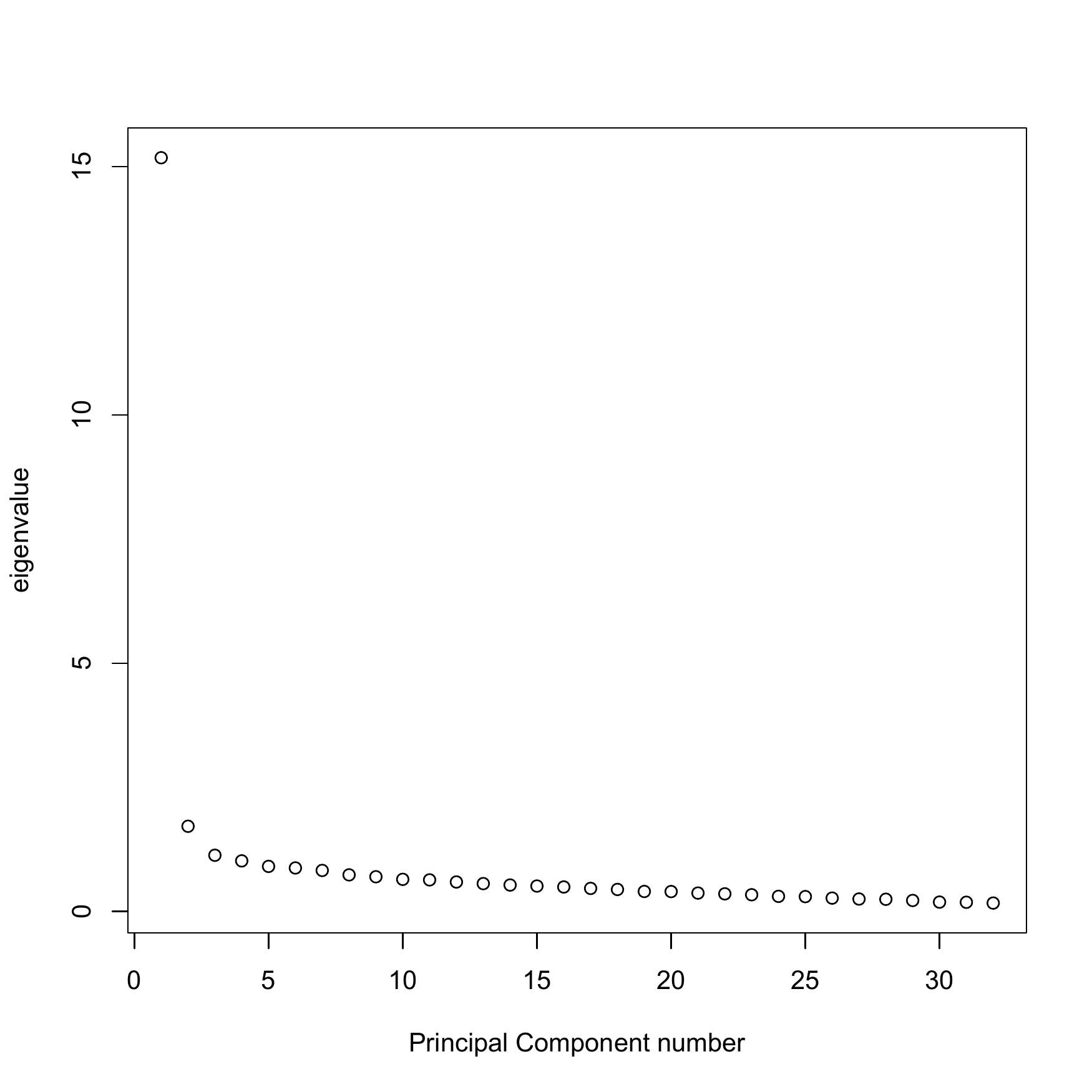}
\end{subfigure}
\begin{subfigure}{0.5\textwidth}
  \centering
  \includegraphics[width=.9\linewidth, height=.25\textheight]{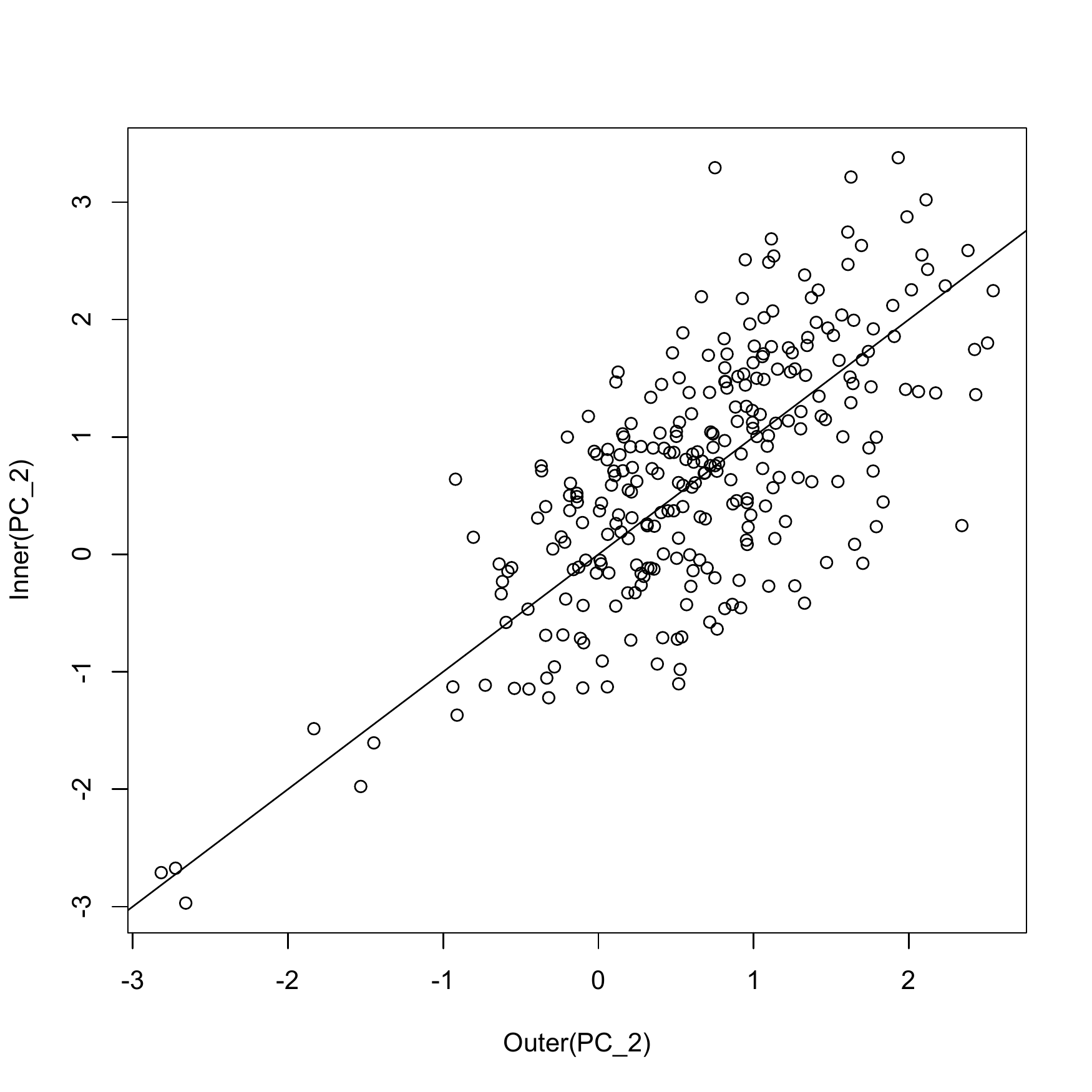}
\end{subfigure}
\\
\begin{subfigure}{0.5\textwidth}
  \centering
  \includegraphics[width=.9\linewidth, height=.25\textheight]{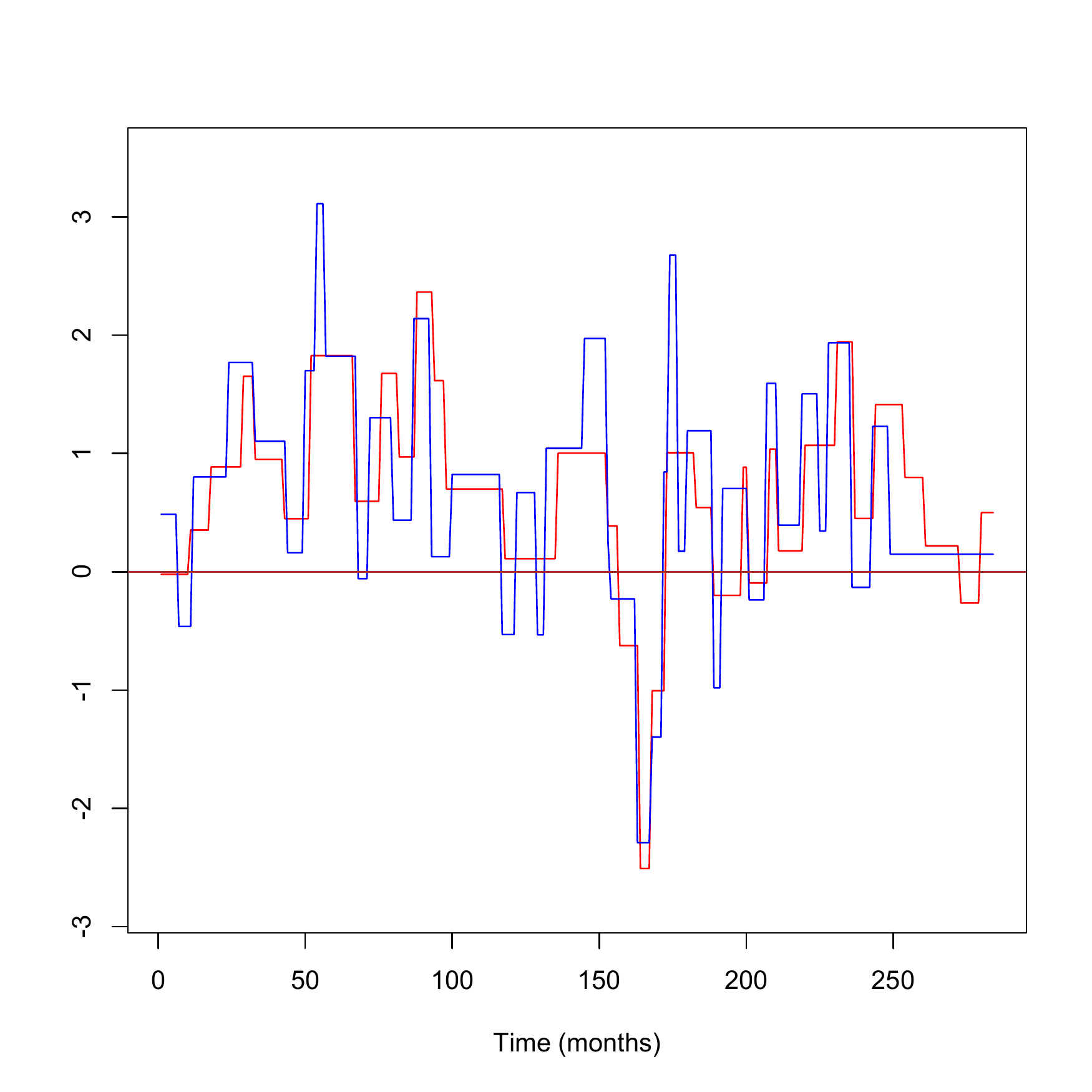}
\end{subfigure}
\begin{subfigure}{0.5\textwidth}
  \centering
  \includegraphics[width=.9\linewidth, height=.25\textheight]{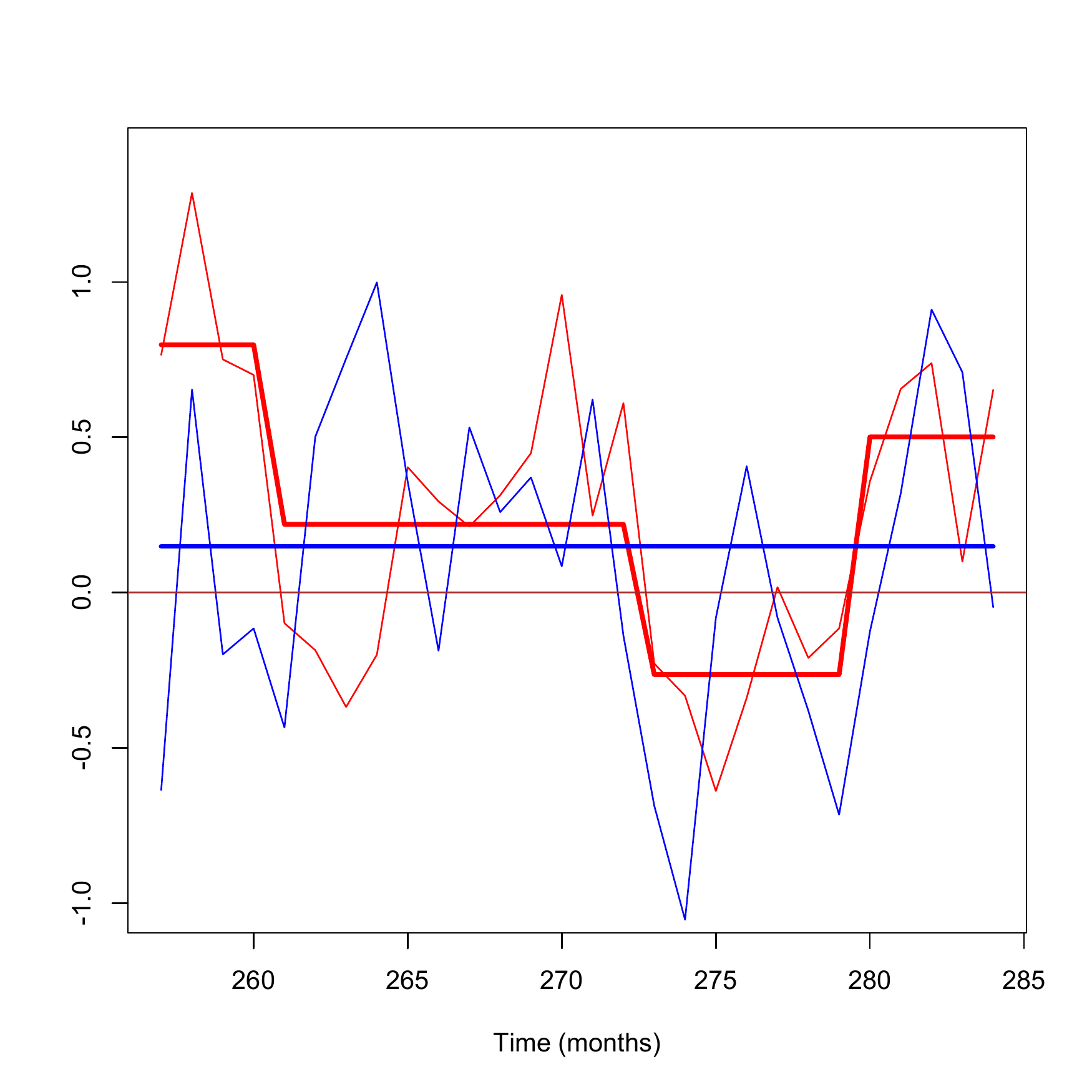}
\end{subfigure}
\\
\begin{subfigure}{0.5\textwidth}
  \centering
  \includegraphics[width=.9\linewidth, height=.25\textheight]{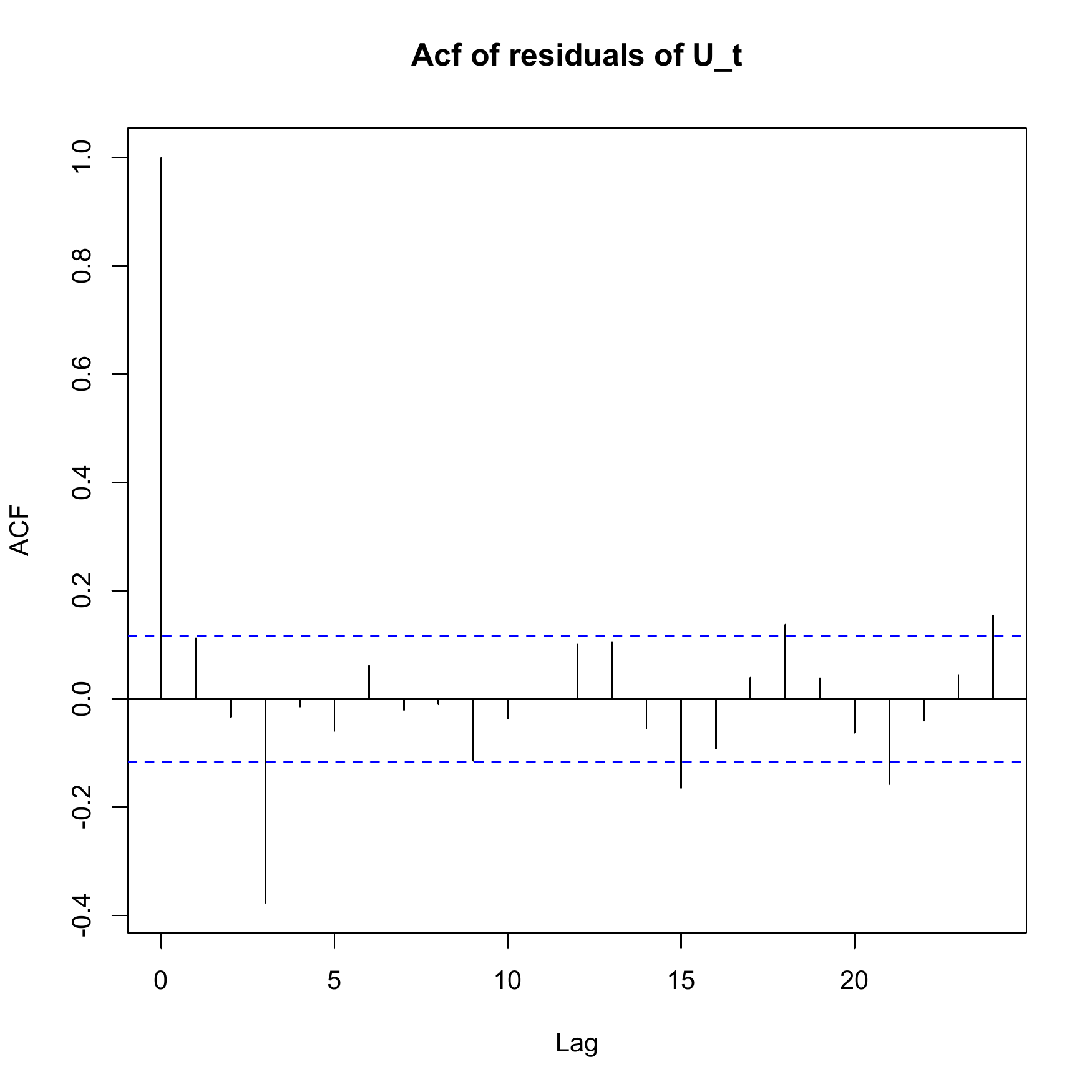}
\end{subfigure}
\begin{subfigure}{0.5\textwidth}
  \centering
  \includegraphics[width=.9\linewidth, height=.25\textheight]{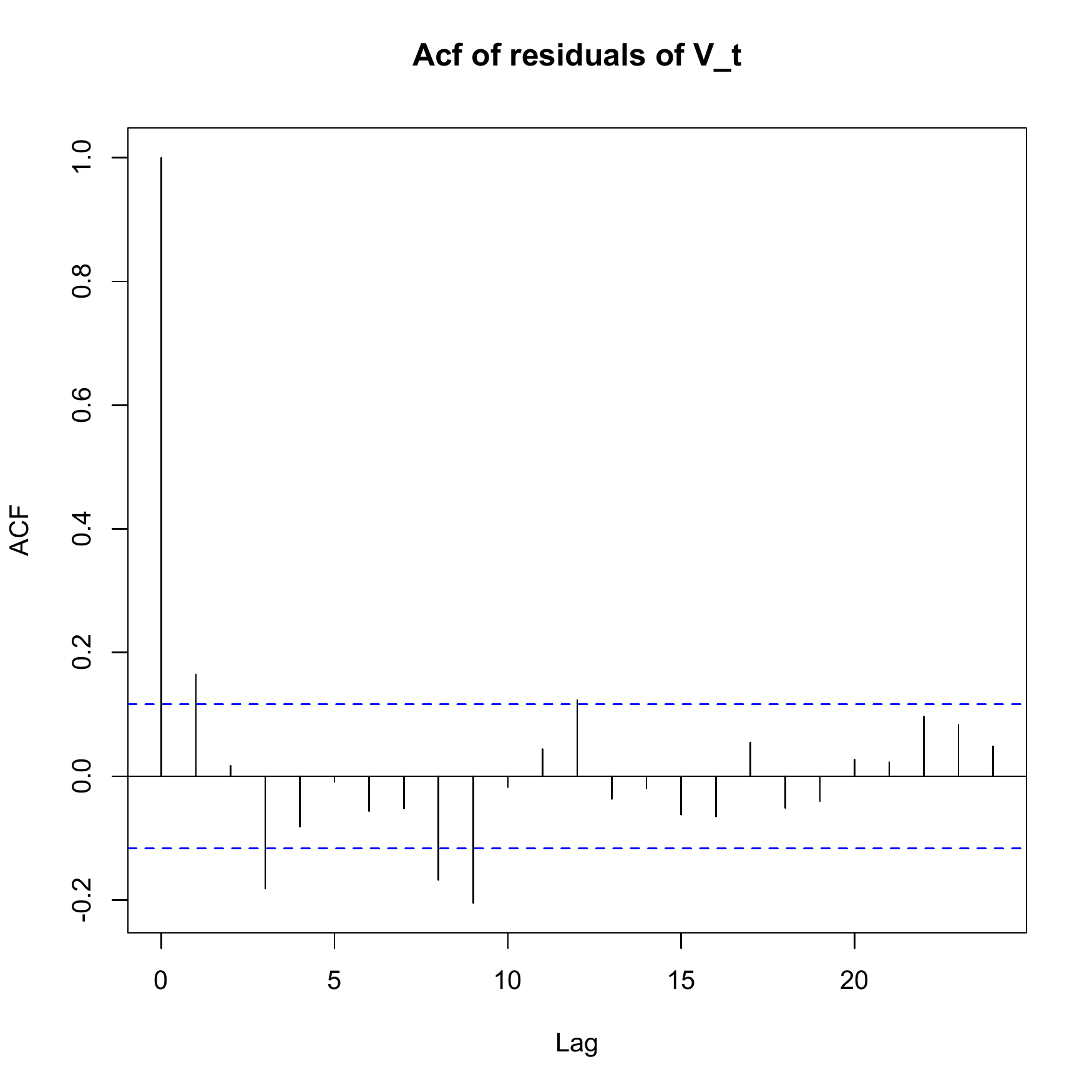}
\end{subfigure}
\caption{Top left: scree plot for the PCA of the (scaled) $X_t$; top right: scatter plot of $U_t$ versus $V_t$; middle left:
$\hat{E}(U_t)$ (blue) and $\hat{E}(V_t)$ (red); middle right: $U_t$ (thin blue), $\hat{E}(U_t)$ (thick blue),
$V_t$ (thin red), $\hat{E}(V_t)$ (thick red) [note: the time ranges in the two middle row plots are different];
bottom left: sample acf of $U_t - \hat{E}(U_t)$; bottom right:
sample acf of $V_t - \hat{E}(V_t)$. See Section \ref{sec:ukhpi} for details.}
\label{fig:london}
\end{figure}

As a preliminary dimension-reduction step, we run the PCA on a version of $X_t$ scaled so that the empirical
second moment of each of the 32 components is one; the scaling appears necessary due to the fact that the empirical
variances of $X_t^{(i)}$ range from 0.97 (for Bromley) to 9.02 (for Kensington and Chelsea). The scree plot is
shown in the top left plot of Figure \ref{fig:london} and appears to offer visual evidence that the 2nd principal component may be of importance
in explaining some of the contemporaneous relationship between the components of $X_t$ beyond the contemporaneous mean
(the 1st PC is close
to the contemporaneous average of $X_t^{(i)}$ and we do not discuss it here).

\begin{table}[h]
\centering
\begin{tabular}{|c|p{60mm}|c|}
  \hline
Indexing set & List of boroughs & Notation \\
\hline
${\mathcal U}$ &
Barnet,
Camden,
City of Westminster,
Ealing,
Hackney,
Hammersmith and Fulham,
Haringey,
Hounslow,
Islington,
Kensington and Chelsea,
Kingston upon Thames,
Lambeth,
Merton,
Richmond upon Thames,
Southwark,
Wandsworth
&
Inner($PC_2$)\\
\hline
${\mathcal V}$ &
Barking and Dagenham,
Bexley,
Brent,
Bromley,
Croydon,
Enfield,
Greenwich,
Harrow,
Havering,
Hillingdon,
Lewisham,
Newham,
Redbridge,
Sutton,
Tower Hamlets,
Waltham Forest &
Outer($PC_2$)\\
\hline
\end{tabular}
\caption{The memberships of the sets ${\mathcal U}$ (Inner($PC_2$) boroughs) and ${\mathcal V}$ (Outer($PC_2$) boroughs). See
Section \ref{sec:ukhpi} for details.
\label{tab4}}
\end{table}

\begin{figure}[h]
  \centering
  \includegraphics[width=.8\linewidth, height=.52\textheight]{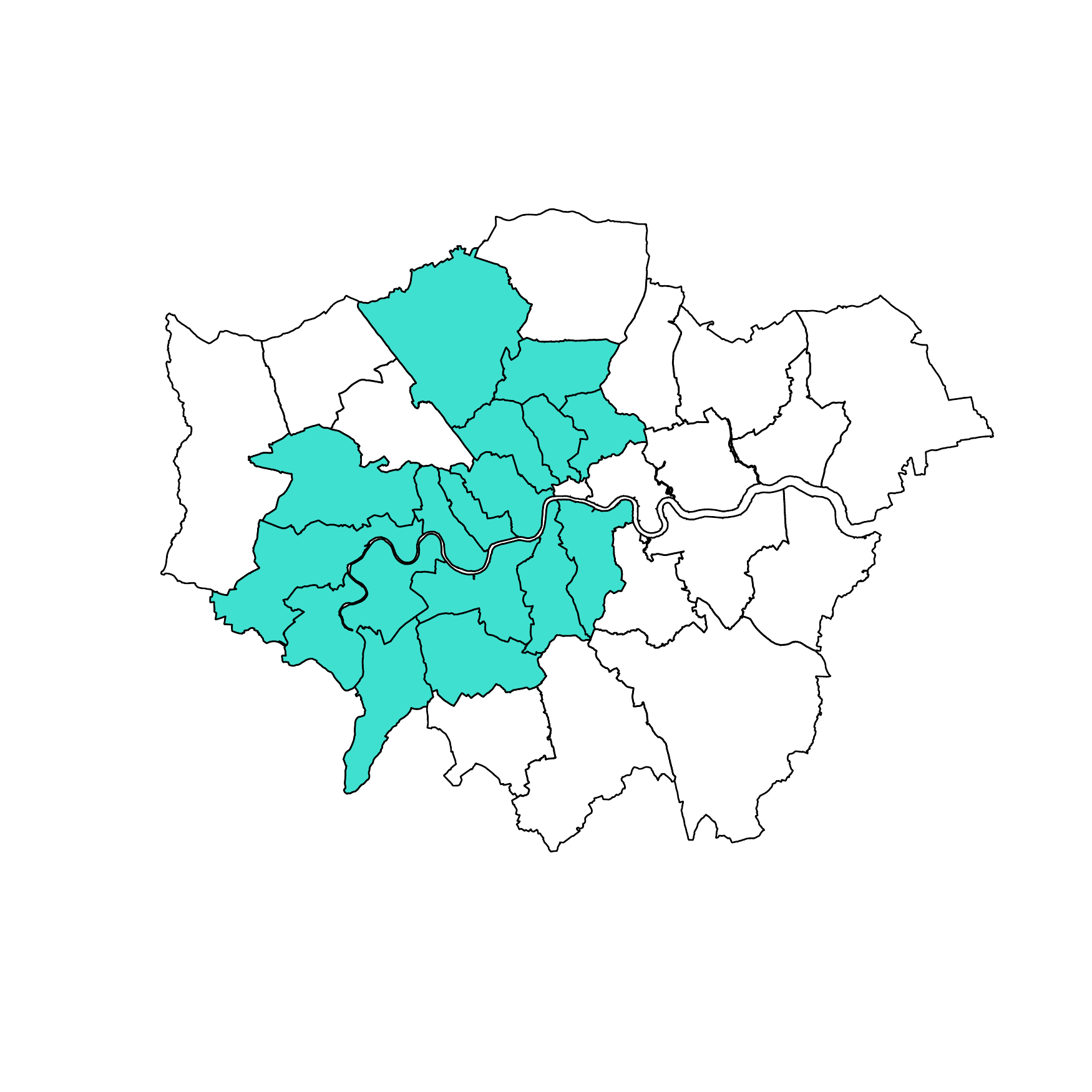}
\caption{London boroughs. Turquoise: the Inner($PC_2$) boroughs (included in set ${\mathcal U}$); white: the Outer($PC_2$) boroughs (included in set
${\mathcal V}$). See Table \ref{tab4} for a list, and Section \ref{sec:ukhpi} for details.}
\label{fig:boroughs}
\end{figure}

Let $\omega_i^{(2)}$ denote the loading of $X_t^{(i)}$ in the 2nd principal component of $X_t$.
The 2nd PC appears to furnish an interesting and interpretable
separation of $X_t$ according to the sign of $\omega_i^{(2)}$. Let 
${\mathcal U}, {\mathcal V}$ be two complementary subsets of 
${\mathcal J} = \{1, 2, \ldots, 32\}$ corresponding to the opposing signs of
$\omega_i^{(2)}$; that is, ${\mathcal U} \cup {\mathcal V} = {\mathcal J}$;
${\mathcal U} \cap {\mathcal V} = \emptyset$ and
$\forall\,i\in {\mathcal U}\,\,\,\forall\,j\in {\mathcal V}\,\,\, \omega_i^{(2)} \omega_j^{(2)} < 0$.
The assignment of boroughs to sets 
${\mathcal U}$ and ${\mathcal V}$ is in Table \ref{tab4}, and it is visualised in Figure \ref{fig:boroughs}.
It is interesting to observe that the 16 boroughs forming part of set ${\mathcal U}$, which we refer to
as ``Inner($PC_2$)" below, overlap very strongly with London's most expensive boroughs by average house
price as of September 2018 (source: {\footnotesize \url{https://www.gov.uk/government/publications/uk-house-price-index-england-september-2018/uk-house-price-index-england-september-2018}}). We refer to the boroughs in set ${\mathcal V}$ as ``Outer($PC_2$)".
The Inner($PC_2$) form a spatially contiguous area; the Outer($PC_2$) boroughs are made up of
two such areas. In summary, the 2nd PC appears to act as a separator between the relatively expensive and inexpensive London boroughs.

We form separate contemporaneous averages of $X_t$ over the sets ${\mathcal U}$ and ${\mathcal V}$,
i.e. corresponding to the Inner($PC_2$) and Outer($PC_2$) boroughs, respectively. We denote
\[
U_t = \frac{1}{16} \sum_{i \in {\mathcal U}} X_t^{(i)}, \qquad V_t = \frac{1}{16} \sum_{i \in {\mathcal V}} X_t^{(i)}.
\]
The scatter plot of $U_t$ versus $V_t$ is shown in the top right plot of Figure \ref{fig:london}.
We model both $U_t$ and $V_t$ as piecewise-constant signals observed with noise. 
We are interested in a comparative analysis of the estimated change-point locations in 
$E(U_t)$ and $E(V_t)$ and the behaviour of the estimates of $E(U_t)$ and $E(V_t)$ between the
estimated change-points. To this
end, we use the WBS2.SDLL(0.9) method.
The resulting piecewise-constant estimates of 
$E(U_t)$ and $E(V_t)$, denoted, respectively, by $\hat{E}(U_t)$ and $\hat{E}(V_t)$, are shown in Figure \ref{fig:uv} (together
with the data $U_t$ and $V_t$, respectively) and 
in the middle left plot of Figure \ref{fig:london} (without the data).
Both $\hat{E}(U_t)$ and $\hat{E}(V_t)$ contain frequent change-points (38 and 33, respectively, in a sample of length
284), which justifies the use of WBS2.SDLL. The bottom plots in Figure \ref{fig:london} show the sample autocorrelation
functions of the empirical residuals $U_t - \hat{E}(U_t)$ and $V_t - \hat{E}(V_t)$ from the respective piecewise-constant fits.
The estimated residuals display only a limited degree of autocorrelation. The piecewise-constant fits also remove practically all
contemporaneous correlation between $U_t$ and $V_t$: the sample correlation between
$U_t$ and $V_t$ is $0.7$, while the sample correlation between
$U_t - \hat{E}(U_t)$ and $V_t - \hat{E}(V_t)$ is 0.06. It is tempting to conclude that the 
piecewise-constant modelling of $E(U_t)$ and $E(V_t)$ with frequent change-points leads to reasonable
goodness-of-fit.

The British mainstream press and professional bodies regularly comment on the perceived link between the outcome of the 
2016 UK European Union membership referendum and the evolution of house prices in London and elsewhere in the UK, see e.g.
{\footnotesize \url{https://www.theguardian.com/business/2018/dec/13/uk-property-market-at-weakest-since-2012-as-brexit-takes-toll-rics}},
amongst a large amount of material on this topic.
We investigate the behaviour of $\hat{E}(U_t)$ and $\hat{E}(V_t)$ between June 2016, the month in which the referendum was held, and
September 2018; this corresponds to the time range $t = 257, \ldots, 284$ in our data. The corresponding illustration is in the middle
right plot of Figure \ref{fig:london}. It is interesting to observe that $\hat{E}(U_t)$, the estimated average monthly percentage change in
price for the Inner($PC_2$) boroughs, contains no change-points over this time period (and has the numerical value of 0.15), while 
$\hat{E}(V_t)$, the same quantity for the Outer($PC_2$) boroughs, shows three change-points which take it from the positive
value of 0.8 to the negative value of $-0.26$, before a rebound which takes it to the level of 0.5. These three most recent change-points in 
$\hat{E}(V_t)$ occur in October 2016 (the level changes from 0.8 to 0.22), October 2017 (from 0.22 to $-0.26$) and May 2018 (from $-0.26$ to 0.5).
We mention that the larger number of change-points in $\hat{E}(V_t)$ than in $\hat{E}(U_t)$ over this particular time period
is in contrast to the fact that it is $\hat{E}(U_t)$ that has more change-points in total (38 to $\hat{E}(V_t)$'s 33). We conclude that
the average house price growth in the Inner($PC_2$) and Outer($PC_2$) boroughs appears to have followed different paths
in the time period since the referendum: the average growth in the Inner($PC_2$) boroughs has remained relatively slow but
constant, while the average growth in the Outer($PC_2$) boroughs had moved from a relatively high value into the negative territory 
before recovering considerably in the recent months.

\section{Discussion}

We emphasise the modular character of the proposed WBS2.SDLL change-point detection methodology:
one ingredient is the WBS2 solution path, and the other is the SDLL model selection criterion. The two parts
have been designed with each other in mind, but it is in principle possible to use them separately; that is, to use
WBS2 with an alternative model selection criterion, and to attempt to adapt the SDLL model selection principle 
to other settings, including ones unrelated to change-point detection.

This paper introduces WBS2 and SDLL in the possibly simplest change-point detection setting, in which a
one-dimensional signal is observed in i.i.d. Gaussian noise; part of this paper's message that even this simplest
framework was traditionally challenging for the state of the art. We believe that WBS2.SDLL fills an important gap in the
literature by providing a solution to this canonical problem in frequent-change-point scenarios. Just as the WBS 
procedure was extended from the canonical signal + i.i.d. Gaussian noise setting to univariate time series \citep{kf17}
and high-dimensional panel data \citep{ws18}, we envisage that WBS2's and SDLL's simplicity means that they are
both extendible to these and other stochastic models.

\appendix

\section{Proofs}

{\bf Proof of Theorem \ref{th:solpath}.}

{\em Part (i).} The statement is trivially true if $T = 1$. Suppose, inductively, that it is true for data lengths $1, \ldots, T-1$.
For an input data sequence of length $T$, the first addition to the solution path breaks the domain of operation into two,
of lengths $b_{m_0}$ and $T-b_{m_0}$. Therefore by the inductive hypothesis, the length of the solution path for the
input data will be $1 + (b_{m-0} - 1) + (T-b_{m_0} - 1) = T-1$, which completes the proof of part (i).

\vspace{10pt}

{\em Part (ii).} The computation of the CUSUM statistic for all $b$ in formula (\ref{eq:ip}) is of computational order $O(e-s)$.
Therefore the addition of elements to the solution path from all sub-domains within a single scale of operation is
of computational order $O(\tilde{M}T)$. If the scale has reached $J$, there is nothing to do and the procedure has stopped.
Therefore the overall computational complexity before the sorting step is of order $O(\tilde{M}JT)$. The sorting can be performed
in computational time $O(T \log\,T)$, which never dominates $O(\tilde{M}JT)$ as the smallest possible $J$ is of computational
order $O(\log\,T)$, which is straightforward to see from its definition. This completes the proof of part (ii).

\vspace{10pt}

{\em Part (iii).} We first set the probabilistic framework in which we analyse the behaviour of the WBS2 solution path algorithm.
With the change-point locations denoted by $\eta_1, \ldots, \eta_N$ (with the additional notation $\eta_0 = 1$,
$\eta_{N+1} = T+1$), we define the intervals
\[
I_{i,j}^{k,l} = [\max(1, \eta_i + k), \min(T, \eta_j + l)],
\]
where $k,l \in \mathbb{Z} \cap [- C_1(\Delta)(\underline{f}_T)^{-2}\log\,T, C_1(\Delta)(\underline{f}_T)^{-2}\log\,T]$
for each $1 \le i+1 < j \le {N+1}$.
Suppose that on each interval
$I_{i,j}^{k,l}$, $\tilde{M}$ intervals $\{[s_m, e_m]\}_{m=1}^{\tilde{M}}$ have been drawn, with the start- and end-points having been
drawn uniformly with replacement from $I_{i,j}^{k,l}$ (if $\tilde{M} > |I_{i,j}^{k,l}|(|I_{i,j}^{k,l}| - 1)/2$, then we understand
$\{[s_m, e_m]\}_{m=1}^{\tilde{M}}$ to contain all possible subintervals of $I_{i,j}^{k,l}$). Note we do not reflect the (stochastic)
dependence of $s_m, e_m$ on $i, j, k, l$ so as not to over-complicate the notation. As in the proof of Theorem 3.2 in
\cite{f14a}, we define intervals ${\mathcal I}_r$ between
change-points in such a way that their lengths are at least of order $O(T)$, and they are separated from the change-points 
also by distances at least of order $O(T)$. To fix ideas, define ${\mathcal I}_r = [\eta_{r-1} + \frac{1}{3}(\eta_r - \eta_{r-1}), 
\eta_{r-1} + \frac{2}{3}(\eta_r - \eta_{r-1})]$, $r = 1, \ldots, N+1$. For each interval $I_{i,j}^{k,l}$, we are interested in the following event
\[
A_{i,j}^{k,l} = \{  \exists_{m_0\in\{1, \ldots, \tilde{M}\}} \,\, \exists_{r\in\{i+1,\ldots,j-1\}}\,\, (s_{m_0}, e_{m_0}) \in {\mathcal I}_r \times {\mathcal I}_{r+1} \}.
\]
Note that
\begin{eqnarray*}
P\{(A_{i,j}^{k,l})^c\} & \le & \prod_{m=1}^{\tilde{M}} P\left\{  (s_m, e_m) \not\in \bigcup_{r=i+1}^{j-1} {\mathcal I}_r \times {\mathcal I}_{r+1}  \right\}\\
& \le & \prod_{m=1}^{\tilde{M}} \max_{r\in\{i+1,\ldots,j-1\}} (1 - P\{(s_m, e_m) \in {\mathcal I}_r \times {\mathcal I}_{r+1}\} ) \le (1 - \delta^2/9)^{\tilde{M}},
\end{eqnarray*}
and hence
\[
P\left(  \bigcap_{i,j,k,l} A_{i,j}^{k,l}  \right) \ge 1 - \sum_{i,j,k,l} P\{(A_{i,j}^{k,l})^c\} \ge 1 - \frac{1}{2}N(N+1) (2C_1(\Delta)(\underline{f}_T)^{-2}\log\,T + 1)^2 (1 - \delta^2/9)^{\tilde{M}}.
\]
Define further the following event
\[
{\mathcal D}_{\Delta, T} = \left\{  \forall\,\, 1 \le s \le b < e \le T\qquad  |\tilde{\varepsilon}_{s,e}^b| \le 2 \sigma \{ (1 + \Delta) \log\, T\}^{1/2}   \right\},
\]
where $\tilde{\varepsilon}_{s,e}^b$ is defined as in formula (\ref{eq:ip}) with $\varepsilon_t$ in place of $X_t$.
It is the statement of Lemma A.1 in \cite{f18} that ${\mathcal B}_{\Delta, T}
\subseteq {\mathcal D}_{\Delta, T}$.

The following arguments apply on the set $\bigcap_{i,j,k,l} A_{i,j}^{k,l} \cap {\mathcal D}_{\Delta, T}$ for any fixed $\Delta > 0$. Proceeding exactly
as in the proof of Theorem 3.2 in \cite{f14a}, at the start of the procedure we have $s = 1$, $e = T$, and in view of the fact that we are
on $A_{0,N+1}^{0,0} \cap {\mathcal D}_{\Delta, T}$, the procedure finds a $b_{m_0} \in [s_{m_0}, e_{m_0}] \subseteq [s,e]$
such that
\begin{enumerate}
\item[(a)] there exists an $r \in \{1, 2, \ldots, N\}$ such that $|\eta_r - b_{m_0}| \le C_1(\Delta)(\underline{f}_T)^{-2}\log\,T$, and
\item[(b)] $|\tilde{X}_{s_{m_0}, e_{m_0}}^{b_{m_0}}| \gtrsim T^{1/2} \underline{f}_T$,
\end{enumerate}
where the $\gtrsim$ symbol means ``of the order of or larger''.
Again by arguments identical to those in the proof of Theorem 3.2 in \cite{f14a}, on each subsequent segment containing previously
undetected change-points, we are on one of the sets $A_{i,j}^{k,l} \cap {\mathcal D}_{\Delta, T}$, 
and therefore the procedure again finds a $b_{m_0}$ satisfying properties (a), (b) above for a certain previously 
undetected change-point $\eta_r$, until all change-points have been identified in this way. Once all change-points have been identified, by Lemma
A.5 of \cite{f14a}, all maximisers $b_{m_0}$ of the absolute CUSUM statistics $|\tilde{X}_{s_m, e_m}^b|$ (over $b$) are such
that $|\tilde{X}_{s_{m_0}, e_{m_0}}^{b_{m_0}}| \le C_2(\Delta) \log^{1/2}T$. As $\log^{1/2}T = o(T^{1/2}\underline{f}_T)$, for $T$ large enough, the sorting
of the elements of $\tilde{\mathcal P}$ does not move any elements from the first $N$ to the last $T-1-N$ or vice versa, which completes the
proof of part (iii). $\hfill\square$

\vspace{10pt}

{\bf Proof of Theorem \ref{th:sdll}.}

With the notation as in the statement of the theorem and in the proof of Theorem \ref{th:solpath}, the following arguments
apply on the set $\bigcap_{i,j,k,l} A_{i,j}^{k,l} \cap {\mathcal D}_{\Delta, T} \cap \Theta_{\theta, T}$ for any fixed $\Delta > 0$.
Let $T$ be large enough for the assertion of Theorem \ref{th:solpath} to hold. Further, let $\tilde{C}$ be large enough for the following
inequality to hold
\begin{equation}
\label{eq:dblineq}
\tilde{\zeta}_T = \tilde{C}\hat{\sigma}_T \{2 \log\,T\}^{1/2} > \max( C_2(\Delta)\log^{1/2}T, 2\sigma\{ (1+\Delta) \log\,T  \}^{1/2} )
\end{equation}
(this is always possible on $\Theta_{\theta, T}$). If $N = 0$, then in view of inequality (\ref{eq:dblineq}) and the fact that we
are on ${\mathcal D}_{\Delta, T}$, from the definition of the SDLL algorithm we necessarily have $\hat{N} = 0$. If $N > 0$, then
by part (iii) of Theorem \ref{th:solpath}, we must have $K + 1 \ge N$. If $|\tilde{X}_{s_{K+1}, e_{K+1}}^{b_{K+1}}| > \tilde{\zeta}_T$,
then by inequality (\ref{eq:dblineq}), we have $N = K + 1$ and the SDLL procedure correctly identifies $\hat{N} = K + 1 = N$.
If $|\tilde{X}_{s_{K+1}, e_{K+1}}^{b_{K+1}}| < \tilde{\zeta}_T$, then we necessarily have $N < K+1$, and by part (iii) of Theorem \ref{th:solpath},
we have, for $Z_k = \log|\tilde{X}_{s_k, e_k}^{b_k}| - \log|\tilde{X}_{s_{k+1}, e_{k+1}}^{b_{k+1}}|$,
\begin{eqnarray*}
Z_N & \sim & \log\,T,\\
Z_k & \le & \log\left(  \frac{(1+ \theta) C_2(\Delta)}{ \beta \tilde{C} \sigma 2^{1/2}}    \right)
\quad\text{for}\quad k=N+1, \ldots, K\quad\text{if this range is non-empty}.
\end{eqnarray*}
Therefore, from the definition of the SDLL and by part (iii) of Theorem \ref{th:solpath}, for $T$ large enough, $\hat{N} = N$ will be chosen. This completes the proof of the
Theorem. $\hfill\square$

\bibliographystyle{plainnat}
\bibliography{pio_2}

\end{document}